\documentclass{svproc}
\usepackage{amsmath}
\usepackage{amsfonts}
\usepackage{url}
\usepackage{float}
\usepackage{graphicx}
\usepackage{subcaption}

\begin{document}
\mainmatter

\title{DeepPM: A Deep Learning-based Profit Maximization Approach in Social Networks}

\titlerunning{DeepPM - Deep Learning based Profit Maximization}

\author{Poonam Sharma \and Suman Banerjee}

\authorrunning{Sharma and Banerjee}

\institute{Indian Institute of Technology Jammu,
J \& K-181221, India. \\
\email{\{poonam.sharma,suman.banerjee\}@iitjammu.ac.in}}

\maketitle

\begin{abstract}
The problem of Profit Maximization asks to choose a limited number of influential users from a given social network such that the initial activation of these users maximizes the profit earned at the end of the diffusion process. This problem has a direct impact on viral marketing in social networks. Over the past decade, several traditional methodologies (i.e., non-learning-based, which include approximate solution, heuristic solution, etc.) have been developed, and many of them produce promising results. All these methods require the information diffusion model as input. However, it may not be realistic to consider any particular diffusion model as real-world diffusion scenarios will be much more complex and need not follow the rules for any particular diffusion model. In this paper, we propose a deep learning-based framework to solve the profit maximization problem. Our model makes a latent representation of the seed sets and is able to learn the diversified information diffusion pattern. We also design a noble objective function that can be optimized effectively using the proposed learning-based approach. The proposed model has been evaluated with the real-world datasets, and the results are reported. We compare the effectiveness of the proposed approach with many existing methods and observe that the seed set chosen by the proposed learning-based approach leads to more profit compared to existing methods. 
The whole implementation and the simulation code is available at: https://github.com/PoonamSharma-PY/DeepPM.
\end{abstract}
\section{Introduction} \label{Sec:Intro}
A \emph{Social Network} is an interconnected structure among a group of human agents which is formed for social interactions. Though the study of social networks has been going on for centuries, however due to the advancement of internet technology and hand hold mobile devices leads to generation of huge amount of social network data. Many of them are publicly available in repositories and these datasets are useful to solve many real-life problems. Among many, one of the important phenomenon is the \emph{diffusion of information}. This phenomena says that any user of a social network tend to share the information in the form of social media post with a hope that this will be liked by many neighbors. Many of the neighbors may like the information and share the post further. This process goes on and the hope is that at the end of the diffusion process a significant number of users of the network will be influenced. In practice, the users from which the diffusion process starts is known as the \emph{seed user}. As real-world social networks are formed using rational human agents, the users who have been chosen as seed needs to be incentivized. Due to the budget constraint only a small number of users can be selected as seed user. Hence, the key computational question arises is that given a social network and a positive integer $k$, which nodes should be chosen as seed to maximize the influence. 
\par Commercial companies exploit the information diffusion phenomenon to promote their brands. To execute this process, they choose a limited number of influential users for initial activation such that the profit earned at the end of diffusion process gets maximized. In this problem, we are given with a social network where each user is assigned with a selection cost and benefit value. The selection cost signifies the amount of incentive that needs to be paid if (s)he has been chosen as a seed user. The benefit value associated with a user signifies the amount can be earned if the user is influenced. A budget has been allocated for the seed set selection process. The goal of the profit maximization problem is to select a limited number of seed nodes within the budget so that the profit earned by the chosen seed set is maximized. Zhu \textit{et al.} first bridged influence and profit maximization in \cite{Zhu2013InfluenceAP} by proposing the \textit{Balanced Influence and Profit (BIP)} problem. Tang \textit{et al.} \cite{8241389} developed approximation algorithms using modified greedy and local search with guarantees for pricing and influence spread. Du \textit{et al.} \cite{10.1007/s10878-021-00774-6} introduced the Marginal Increment–based Prune and Search (MPS) algorithm under profit-aware constraints. Shi \textit{et al.} \cite{SHI202112} defined a \textit{Profit Maximization (PM) problem for social advertising with multiple competing clients}, each providing a budget and an influence requirement. Zhang \textit{et al.} \cite{7524470} and Chen \textit{et al.} \cite{CHEN202036} studied the \textit{Profit Maximization with Multiple Adoptions (PM$^2$A)} problem, where limited seeds must be allocated across products with varying costs and profits. Zhu \textit{et al.} later formulated the \textit{Competitive Profit Maximization (CPro)} problem \cite{8485904}, where a host allocates seeds to competing companies to maximize commission-based profit. Xu \textit{et al.} \cite{XU201213009} proposed a full profit-oriented customer-targeting pipeline incorporating positive and negative influences. Yang \textit{et al.} introduced the \textit{Maximizing Activity Profit (MAP)} problem \cite{8630956}, where profits arise only when groups jointly participate, modeled via hyperedges. Zhu \textit{et al.} further proposed the \textit{Group Profit Maximization (GPM)} problem \cite{10.1007/978-3-030-34980-6_13}, where a group activates once a threshold fraction of members become active.

In this paper, we make the following contributions: 
\begin{itemize}
    \item We study the Profit Maximization Problem under a deep learning framework, and formally, refer to it as \textbf{DeepPM}.
    \item  The proposed model has been evaluated with real-world social network datasets, and the results have been compared with the existing methods from the literature.
\end{itemize}
The rest of the paper has been organized as follows. Section \ref{Sec:PPD} describes the required background concepts and defines our problem formally. Section \ref{Sec:Solution} describes the proposed model with detailed architecture. The experimental evaluation of the proposed model has been described in Section \ref{Sec:Experiments}. Finally, Section \ref{Sec:Conclusion} concludes our study.

\section{Preliminaries and Problem Definition} \label{Sec:PPD}
\subsection{Social Networks and Information Diffusion}
As mentioned previously, a social network is an interconnected structure among a group of human agents which is formed for social interactions. A social network is represented by a weighted, directed graph $\mathcal{G}(\mathcal{V}, \mathcal{E}, \mathcal{P})$. Here, the vertex set $\mathcal{V}$ represents the set of users and the edge set $\mathcal{E}$ represents the set of social ties among the users. $\mathcal{P}$ represents the edge weight function that maps each edge to its corresponding probability value, i.e., $\mathcal{P}: \mathcal{E} \longrightarrow (0,1]$. For any edge $(u_iu_j) \in \mathcal{E}$, its probability is denoted by $\mathcal{P}(u_iu_j)$. 
\par As mentioned earlier, among many, one of the important phenomenon of social networks is the diffusion of information. We consider that the diffusion of information in the network is happening by the rule of some diffusion model (say $\mathcal{M}$). The set of nodes from which the diffusion process starts is called as seed nodes. At the end of diffusion process, a subset of the nodes will be influenced which we call as the influence of the seed set $\mathcal{S}$ under the diffusion model $\mathcal{M}$. It is denoted by $\sigma_{\mathcal{M}}(\mathcal{S})$. Here, $\sigma$ denotes the social influence function that maps each subset of the user to the corresponding influence value, i.e., $\sigma: 2^{\mathcal{V}} \longrightarrow \mathbb{R}_{0}^{+}$ with $\sigma(\emptyset)=0$. In many real-life situations, the goal is to choose a limited number of influential nodes to maximize the influence. This problem is called as the Social Influential Maximization Problem. Given a social network $\mathcal{G}(\mathcal{V}, \mathcal{E}, \mathcal{P})$ and a positive integer $k$, this problem asks to choose $k$ many users as seed to maximize the influence in the network.
\subsection{Profit Maximization Problem}
To study the Profit Maximization Problem, we consider that the users of the network are associated with two positive real numbers. The first one is referred to as the selection cost which signifies that the amount of incentive needs to be paid to that user. This is mathematically formalized as the cost function $\mathcal{C}$ defined as $\mathcal{C}: \mathcal{V} \longrightarrow \mathbb{R}^{+}$. For any user $u_i \in \mathcal{V}$, the selection cost of $u_i$ is denoted by $\mathcal{C}(u_i)$. For any subset of users $\mathcal{S} \subseteq \mathcal{V}$, the total selection cost is denoted by $\mathcal{C}(\mathcal{S})$ and defined as $\mathcal{C}(\mathcal{S})=\underset{u \in \mathcal{S}}{\sum} \ \mathcal{C}(u)$. The second one is referred to as the benefit value which signifies the amount of benefit that can be earned if the user is influenced. This has been formally characterized by a benefit function $b: \mathcal{V} \longrightarrow \mathbb{R}^{+}$. For any user $u_i$, the benefit value associated with this user is denoted by $b(u_i)$. 
\par Now, given a seed set $\mathcal{S}$, the set of nodes that are influenced by $\mathcal{S}$ is denoted by $I(\mathcal{S})$. The benefit earned by the seed set $\mathcal{S}$ is denoted by $\beta(\mathcal{S})$ and defined as the sum of the benefit values of the users who are influenced by the seed set $\mathcal{S}$, i.e., $\beta(\mathcal{S})=\underset{u_i \in I(\mathcal{S})}{\sum} \ b(u_i)$. Here, $\beta()$ is the earned benefit function that maps each subset of nodes to the corresponding earned benefit value. Hence, $\beta: 2^{\mathcal{V}} \longrightarrow \mathbb{R}_{0}^{+}$ with $\beta(\emptyset)=0$. If we subtract the cost of the seed set then we get the profit of the seed set. For any seed set $\mathcal{S}$, the profit earned by $\mathcal{S}$ is denoted by $\Phi(\mathcal{S})$ and defined as $\Phi(\mathcal{S})=\beta(\mathcal{S}) - \mathcal{C}(\mathcal{S})$. The Problem of Profit Maximization asks to choose a subset of the nodes as seed for initial activation such that the selection cost of the set is less than equal to the allocated budget to maximize the profit. The Profit Maximization Problem can be presented as a discrete optimization problem as shown in Equation (\ref{Eq:Problem}).
\begin{equation} \label{Eq:Problem}
    \mathcal{S}^{OPT} \longleftarrow \underset{\mathcal{S} \subseteq \mathcal{V} \text{ and }\mathcal{C}(\mathcal{S}) \leq B}{argmax} \ \Phi(\mathcal{S})
\end{equation}
Here, $\mathcal{S}^{OPT}$ denotes the optimal seed set for allocated budget $B$.

\section{Proposed Model} \label{Sec:Solution}
\begin{figure}[t] 
  \centering
  \includegraphics[width=0.9\linewidth]{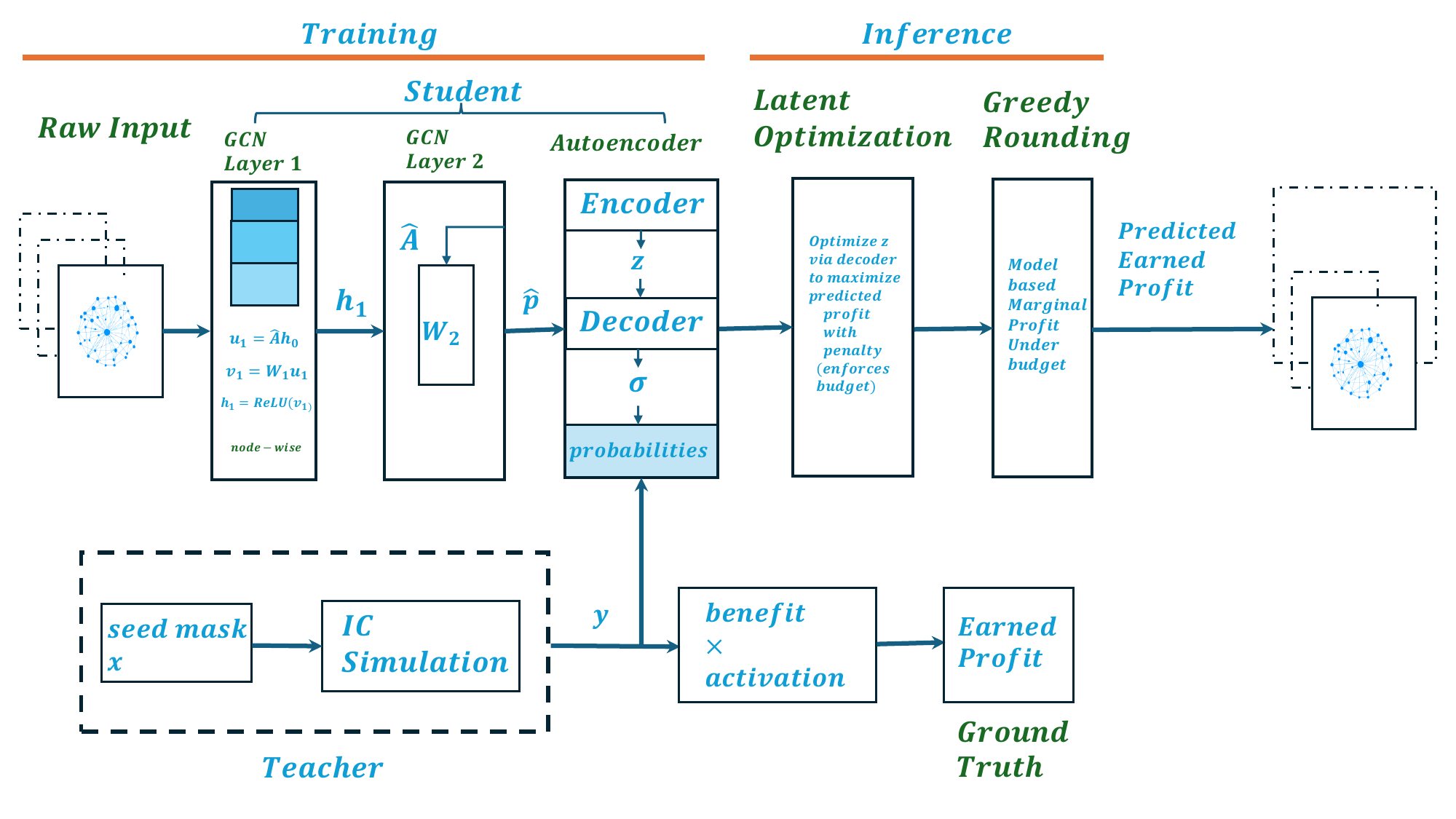}
  \caption{Deep Learning-Based Profit Maximization for Social Networks: DeepPM Model}
  \label{fig:DeepPM}
\end{figure}

In this section, we will understand the working of our deep learning-based model shown in Fig. $\ref{fig:DeepPM}$ for the profit maximization problem. 
For message passing, we use the undirected, unweighted adjacency $\mathcal{A}$, which is taken from $\mathcal{E}$, the degree matrix $D$, and its normalized form.
\begin{equation}
\tilde{\mathcal{A}} = \mathcal{A} + I,\quad D = \operatorname{diag}(\tilde{\mathcal{A}}\mathbf{1}),\quad \hat{\mathcal{A}} = D^{-1/2}\tilde{\mathcal{A}} D^{-1/2}.
\label{eq1:NormalizedEq}
\end{equation}

The above three-part equation construction is the standard symmetric normalization of a graph adjacency used in GCN-based message passing. The symmetric normalization is used to obtain a stable, balanced, and spectrally well-behaved propagation operator for GNN layers.

\subsubsection{Teacher Formulation}
The teacher is a highly faithful diffusion engine that generates ground-truth outcomes. For convenience, we denote by $c \in \mathbb{R}^{|\mathcal{V}|}$ and $b \in \mathbb{R}^{|\mathcal{V}|}$ the vectors whose entries are $c_i = \mathcal{C}(u_i)$ and $b_i = b(u_i)$, respectively. It takes a \emph{budget-feasible seed mask} $x\in\{0,1\}^{|\mathcal{V}|}$ so that $c^\top x\le B$ (it is in the budget limit) indicating which nodes are initially activated, and the \emph{edge diffusion probabilities} $\mathcal{P}$ that govern the chance of transmission along each directed edge.
It then runs the Independent Cascade (IC) process, starting from the seed nodes. Time advances in a discrete manner; whenever a node $u$ becomes newly active, it receives exactly one opportunity to activate each out-neighbor $v$, succeeding independently with probability $\mathcal{P}({uv})$. The process terminates when a full wave produces no new activations. A single roll out returns a stochastic \emph{final activation mask} $y\in\{0,1\}^{\mathcal{|V|}}$, where $y_i=1$ indicates that node $i$ ended up active in that run. We repeat the process many times with the same $(x,\mathcal{P})$, the empirical frequency of $y_i=1$ converges to the \emph{true activation probability} $p_i(x)$:
\[
\mathbb{E}[\,y_i \mid x\,] \;=\; p_i(x),\qquad y_i \mid x \sim \mathrm{Bernoulli}\!\big(p_i(x)\big).
\]



These outcome vectors $y$ are the \emph{teacher labels} that supervise the student. The student model is trained to relate seed masks $x$ to calibrated activation probabilities by aligning its predictions with those of the teacher, who observes the actual outcomes.

\subsubsection{Student Formulation}
The GCN predictor, surrogate, is a lightweight, fully differentiable model that takes (i) a seed mask $x$ (hard $0/1$ or soft $[0,1]$) and (ii) the graph (through its symmetrically normalized adjacency $\hat A$), and returns for every node $i$ a probability $\hat p_i(x)\in(0,1)$ that the node ends up active. 
It is trained to numerically imitate the teacher’s diffusion outcomes while remaining cheap to evaluate and easy to differentiate, which is crucial for profit optimization with gradients. We want $\hat p(\cdot)$ to reflect that influence propagates along edges. Using $\hat A$ degree-normalized adjacency to mix signals implements a stable neighborhood aggregation: one multiplication by $\hat A$ mixes from 1-hop neighbors; two consecutive multiplications mix information up to 2 hops. 
This gives the surrogate a two-hop \emph{receptive field}, often sufficient to rank nodes by marginal influence in many graphs, while keeping the computation $O(|\mathcal{E}|)$.
Given $x\in[0,1]^{\mathcal{|V|}}$,
\begin{align}
h_0 &= x
\label{eq:gcn-h0}\\[-2pt]
&\text{initial seed intensity at each node}\nonumber\\
h_1 &= \mathrm{ReLU}\!\big(W_1\,\hat A\,h_0\big)
\label{eq:gcn-h1}\\[-2pt]
&\text{propagation $\to$ linear mixing $\to$ nonlinearity}\nonumber\\
\hat p(x;\theta) &= \sigma\!\big(W_2\,\hat A\,h_1\big)\in(0,1)^{|\mathcal{V}|}
\label{eq:gcn-prob}\\[-2pt]
&\text{readout after second propagation}\nonumber
\end{align}
with parameters $\theta=\{W_1,W_2\}$.
Intuitively, $W_1$ learns how to \emph{filter} the incoming one-hop evidence of seeding, $\mathrm{ReLU}$ retains positive evidence, the second $\hat A$ pass spreads that evidence one hop further, and $W_2$ converts the aggregated signal into logits; finally, $\sigma(\cdot)$ turns logits into calibrated probabilities. The symmetric normalization $\hat A=D^{-1/2}(A+I)D^{-1/2}$ balances contributions from high- and low-degree nodes and includes self-loops. This prevents high-degree hubs from dominating the averages and keeps repeated aggregations numerically stable (spectral norm $\le 1$). For a fixed input $x$, the teacher produces labels $y_i\in\{0,1\}$ that are Bernoulli with mean $p_i(x)$. The surrogate outputs $\hat p_i(x)\in(0,1)$, which we interpret as the model’s estimate of that Bernoulli mean.

We fit $\theta$ by minimizing the average Bernoulli negative log-likelihood (binary cross-entropy, BCE) against teacher labels:
\begin{equation}
\mathcal{L}_{\text{diff}}(\theta)
= \frac{1}{T}\sum_{t=1}^T\sum_{i=1}^{\mathcal{|V|}} 
\Big(-y^{(t)}_i\log \hat p_i(x^{(t)};\theta) - (1-y^{(t)}_i)\log(1-\hat p_i(x^{(t)};\theta))\Big).
\label{eq:diff-bce}
\end{equation}
For any fixed $x$, the population risk $\mathbb{E}[\mathrm{BCE}(y_i,\hat p_i(x))]$ is minimized at $\hat p_i(x)=p_i(x)$. $T$ is the total number of training samples, which are teacher-student pairs. Thus, with sufficient data and capacity, minimizing Equation (\ref{eq:diff-bce}) yields \emph{calibrated} probabilities in expectation.
The target objective is profit $b^\top p(x)-c^\top x$, but $p(\cdot)$ is intractable to differentiate.
The surrogate provides a smooth proxy $b^\top \hat p(x)-c^\top x$ whose gradients
\[
\nabla_x \big(b^\top \hat p(x)\big) 
= J_{\hat p}(x)^\top b
\]
are readily available by backpropagation, where $J_{\hat p}(x)$ is the Jacobian of $\hat p$ w.r.t.\ $x$.
This enables gradient-based search (e.g., through a seed decoder) and fast marginal-profit evaluations during greedy rounding. Equation (\ref{eq:gcn-prob}) can be read as logistic regression on node features that are \emph{learned} by two rounds of graph filtering: the first round propagates seed signals to neighbors and applies a learned feature transform; the second round smooths and aggregates again before the final logistic readout. This is the minimal graph-aware architecture that still captures neighbor effects and remains computationally lean. The autoencoder learns the low-dimensional description of the seed node patterns seen during training. At inference, we search in its latent space for soft seedings that look realistic and promise high profit under the surrogate. For $x\in\{0,1\}^{\mathcal{|V|}}$,
\begin{equation}
z=\mathrm{enc}(x)\in\mathbb{R}^Z,\qquad \tilde x=\sigma(\mathrm{dec}(z))\in(0,1)^N,
\end{equation}
with binary cross-entropy reconstruction loss
\begin{equation}
\mathcal{L}_{\text{AE}}(\phi)=\frac{1}{T}\sum_{t=1}^T\sum_{i=1}^{\mathcal{|V|}}\Big(-x^{(t)}_i\log \tilde x^{(t)}_i-(1-x^{(t)}_i)\log(1-\tilde x^{(t)}_i)\Big),
\end{equation}
where $\phi$ are encoder/decoder parameters. The decoder's logits score nodes, indicating coordinates that frequently represent seeds found in effective, feasible sets. We train the surrogate and the autoencoder together so that (i) the surrogate matches the teacher’s outcomes, and (ii) the decoder provides a realistic parameterization of seed masks.

\begin{equation}
\mathcal{L}(\theta,\phi)=\lambda_{\text{diff}}\;\mathcal{L}_{\text{diff}}(\theta)+\lambda_{\text{AE}}\;\mathcal{L}_{\text{AE}}(\phi),
\end{equation}
optimized with Adam on mixed mini-batches of $(x^{(t)},y^{(t)})$.

\subsubsection{Student guided Inference}
We first \emph{search} the decoder’s latent space for a soft seed vector that maximizes predicted profit (surrogate benefit minus cost) under a soft budget penalty. Then we \emph{round} to a discrete set by adding items with positive \emph{model-based} marginal profit until no such items remain or the budget would be exceeded.
Let $x_{\text{soft}}(z)=\sigma(\mathrm{dec}(z))$. Define the surrogate profit with a penalty
\begin{equation}
\widehat{\Pi}(z)=b^\top \hat p\!\big(x_{\text{soft}}(z)\big)-c^\top x_{\text{soft}}(z)-\mu\,[c^\top x_{\text{soft}}(z)-B]_+,
\label{eq:latent}
\end{equation}
and maximize $\widehat{\Pi}(z)$ by gradient ascent in $z$ (or equivalently minimize $-\widehat{\Pi}$).

\section{Experimental Evaluation} \label{Sec:Experiments}
In this section, we describe the experimental evaluation of the proposed solution approach. Initially, we describe the datasets used in our experiments.
\subsection{Datasets Used}
In this study, we have used the following datasets in our experiments:
\begin{itemize}
    \item \textbf{Email-Eu-Core} (Euemail) \cite{yin2017local}: Built from email exchanges in a large European research institution; an edge $(u,v)$ exists if $u$ sent $v$ at least one email. 
    \item \textbf{Ego-Facebook} (Facebook) \cite{NIPS2012_7a614fd0} It consists of `circles' from Facebook. Data was collected from survey participants using Facebook app.
    \item \textbf{Wiki-Vote} (Wikivote) \cite{leskovec2010signed,leskovec2010predicting}: Voting data from Wikipedia’s inception to Jan 2008; nodes are users and a directed edge $(u,v)$ means $u$ voted on $v$.
\end{itemize}
The basic statistics of the datasets are reported in Table \ref{tab:Dataset}.
\vspace{-0.7cm}
\begin{table}[h]
	\centering
	\caption{Basic Statistics of the Datasets}
	\label{tab:Dataset}
	\begin{tabular}{|c|c|c|c|c|c|}
	\hline
	\textbf{Dataset} & \textbf{Type of} & \textbf{Number of} & \textbf{Number of} & \textbf{Maximum} & \textbf{Average} \\ 
	\textbf{Name}           & \textbf{Graph}      & \textbf{Nodes} & \textbf{Edges}  & \textbf{Degree} & \textbf{Degree}       \\ \hline
	Euemail & Directed   & 1005    & 25571    &  347    & 33.25 \\ \hline
    Facebook & Undirected &   4039  &   88234    &   1045    & 43.69 \\ \hline
    Wikivote      & Directed      & 7115  & 103689 &  1167  &  29.15  \\ \hline
	\end{tabular}%
	\label{tab:Datasets}
	\end{table}

\subsection{Experiment Setup}
\subsubsection{Influence Probability} In this study, we consider two different influence probability settings, namely a) \emph{Uniform} and b) \emph{Trivalency}. In Uniform setting, every edge of the network has the same influence probability; i.e., $\forall (u,v) \in \mathcal{E}(\mathcal{G})$, $\mathcal{P}_{uv}=p_c$, where $p_c$ is any constant in between $0$ and $1$. We have experimented with the $p_c = 0.1$. In Trivalency setting, the influence probability of every edge is uniformly at random from this set $\{0.1, 0.01, 0.001\}$. 
\subsubsection{Cost and Benefit}
The cost and benefit of a node are assigned uniformly at random from the intervals $[50,100]$ and $[800, 1000]$, respectively.

\subsubsection{Implementation setup}
The experiments were run on a Linux system with an Intel i9, 32 cores at 3.2 GHz, and 64 GiB RAM. All algorithms were implemented in Python 3.13.5, primarily using libraries NetworkX 3.5 and Torch 2.8.

\subsection{Algorithms in our Experimentation}
\subsubsection{Proposed Approach}
\begin{itemize}
    \item \textbf{DeepPM}: DeepPM learns influence probabilities directly from cascades using a GCN-based student–teacher framework and selects seeds by maximizing predicted profit under a given budget.
\end{itemize}
\subsubsection{Baselines} 
In our experimentation, the following methods are involved. 
	\begin{itemize}
    \item \textbf{Stochastic Greedy (StG):}
    Speeds up classical greedy influence maximization by evaluating only a random subset of candidate nodes at each iteration.
	\item \textbf{Simple Greedy Approach (SG):} The nodes that produce the maximum marginal gain are selected iteratively for the seed set within the allocated budget.
	
	\item \textbf{Double Greedy Approach (DG):} 
	Explores the solution space from both ends, starting with an empty set and a full set, and shrinking and growing them simultaneously based on marginal gain.
	
	\item \textbf{Single Discount (SD):} We pick the highest degree node, remove it from the graph, and reduce the degree of each neighbor by one. We repeat this process to add seed nodes to the seed set, within the budget limits.
	
	\item \textbf{Degree Discount (DD):} Same as Single Discount, except that the decrease in the degree of neighbors of a node in the seed set is by $dd_{v} = d_{v} - 2t_{v} - (d_{v} - t_{v})t_{v}p$, as explained in \cite{chen2009efficient}.
	
	\item \textbf{High Degree (HD):} Take the maximum degree node from the graph and continue to select the next highest degree node to include in the seed set within budget limits. 
	
	\item \textbf{High Clustering Coefficient (HC):} The clustering coefficient of each node is computed and sorted in non-increasing order based on its value. The seed nodes are selected within the budget limits. 
	
	\item \textbf{Random:} The seed nodes are selected uniformly at random within the budget. 
	\end{itemize}

\subsection{Research Questions}
In our experiments, we have analyzed the following research questions:
\begin{itemize}

\item \textbf{RQ1. (Impact of Budget on Profit Earned)}: Under an influence probability setting, how the earned profit is changing with the increasing budget.  
\item \textbf{RQ2. (Impact of Budget on Seed Set Size)}: Under an influence probability setting, how the seed set size is changed with the increasing budget.  
\item  \textbf{RQ3. (Computational Time Requirement)}: Under a fixed influence probability setting and a fixed budget, comparison of computational time requirements for different methodologies.
\end{itemize}
\subsection{Experimental Results and Description}

\subsubsection{RQ1. Impact of Budget on Profit Earned}
\begin{figure}[h!]
\centering

\begin{subfigure}{0.32\textwidth}
    \centering
    \includegraphics[width=\linewidth]{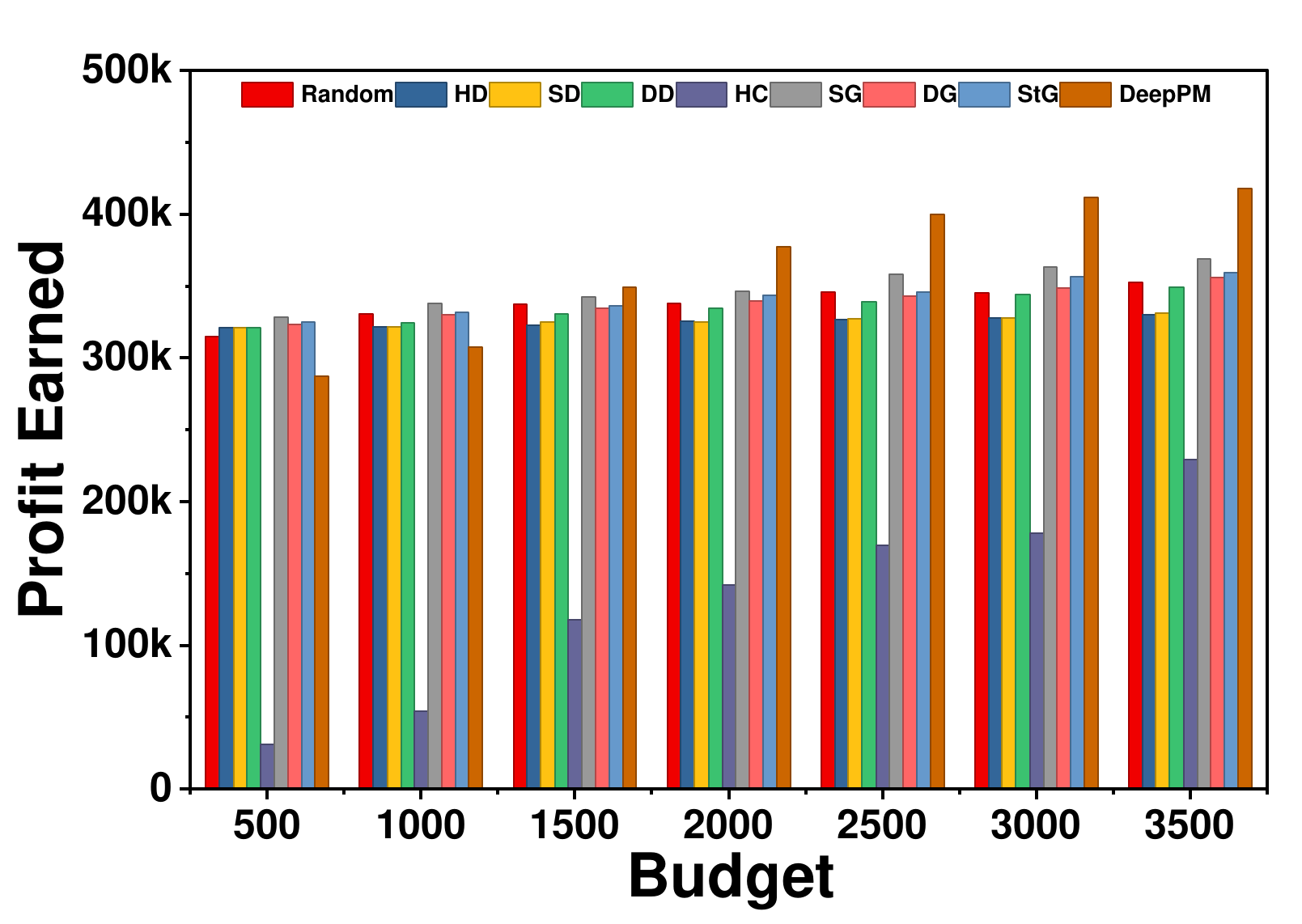}
    \caption{Euemail}
\end{subfigure}
\hfill
\begin{subfigure}{0.32\textwidth}
    \centering
    \includegraphics[width=\linewidth]{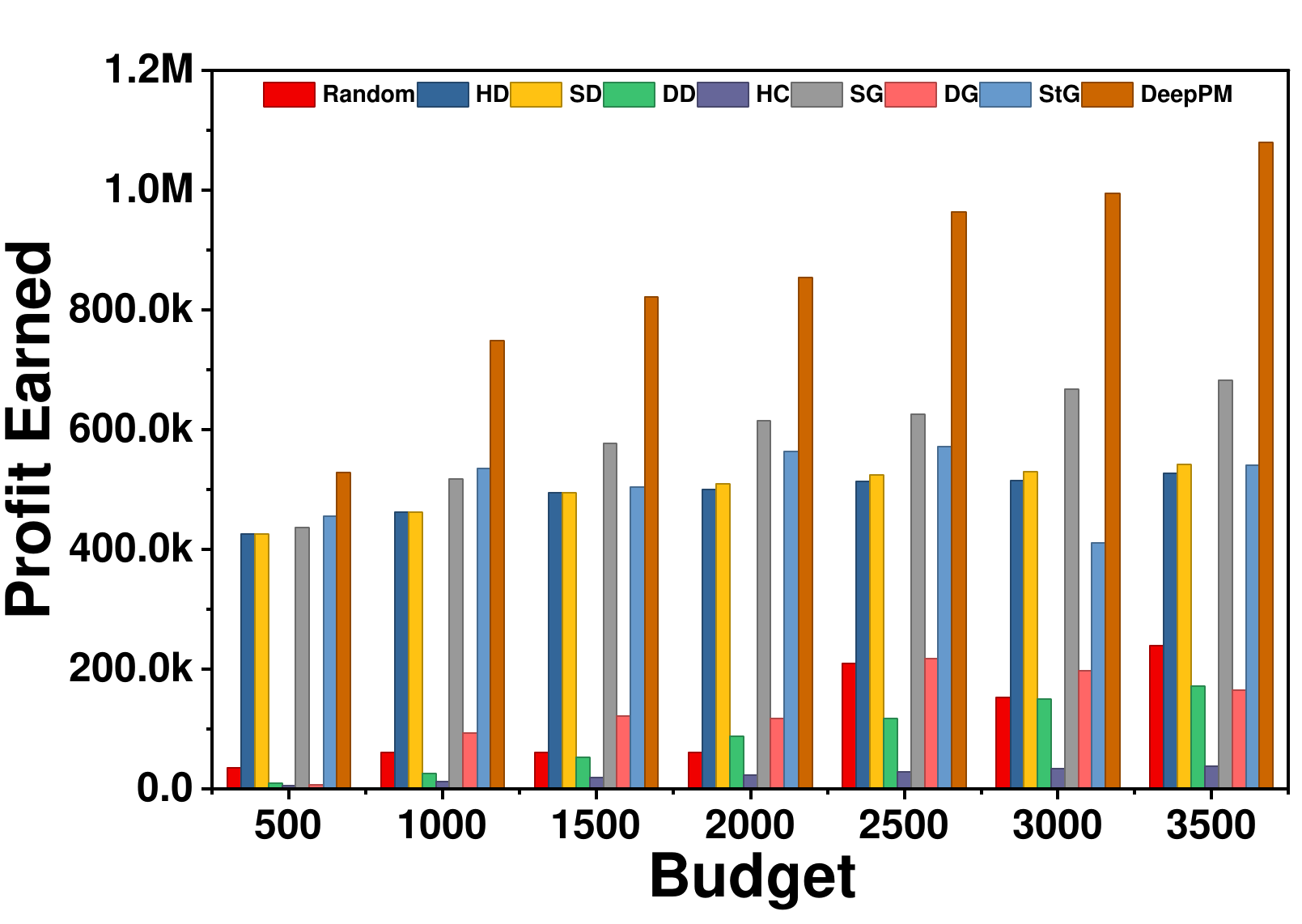}
    \caption{Facebook}
\end{subfigure}
\hfill
\begin{subfigure}{0.32\textwidth}
    \centering
    \includegraphics[width=\linewidth]{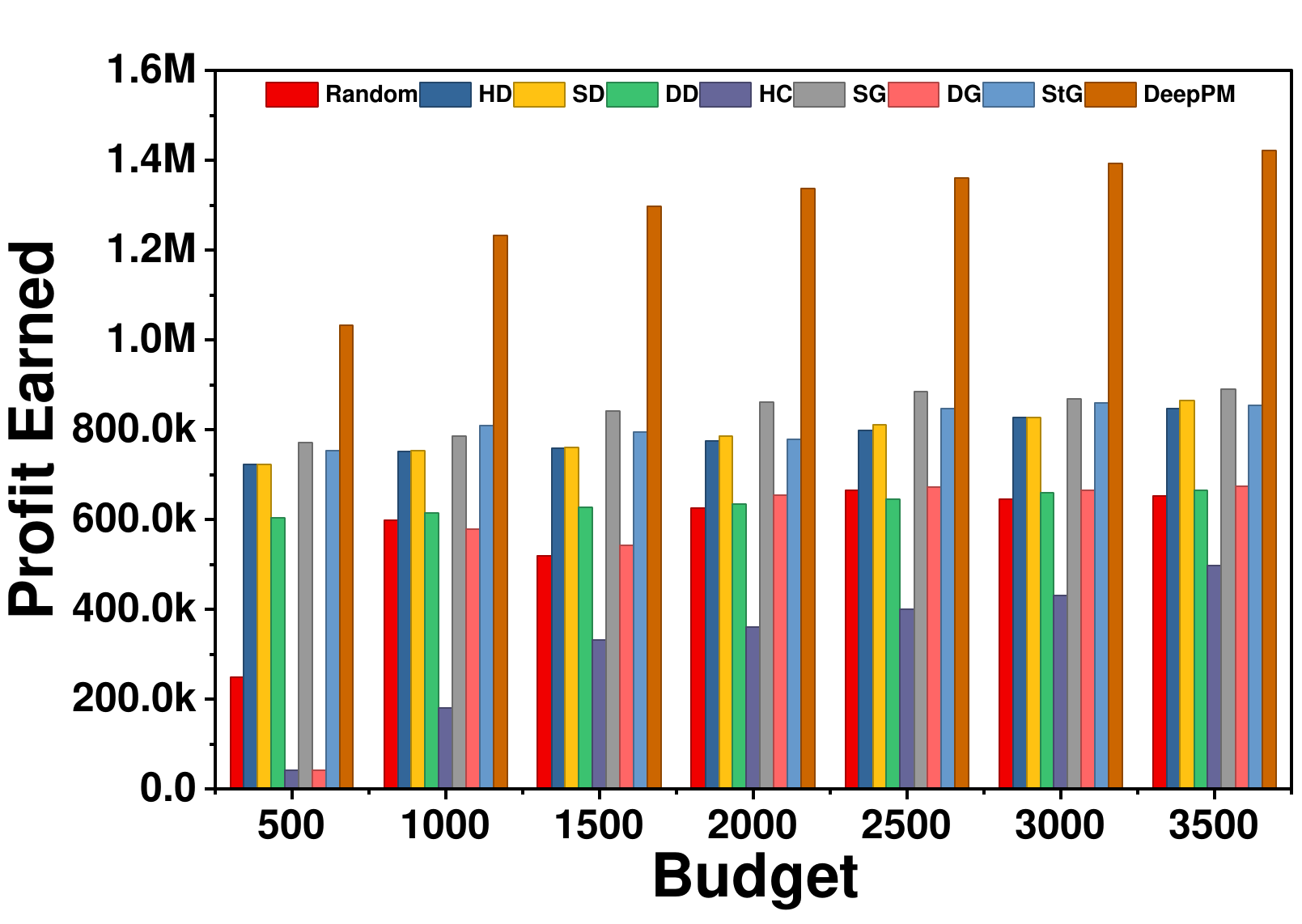}
    \caption{Wikivote}
\end{subfigure}

\vspace{8pt}

\begin{subfigure}{0.32\textwidth}
    \centering
    \includegraphics[width=\linewidth]{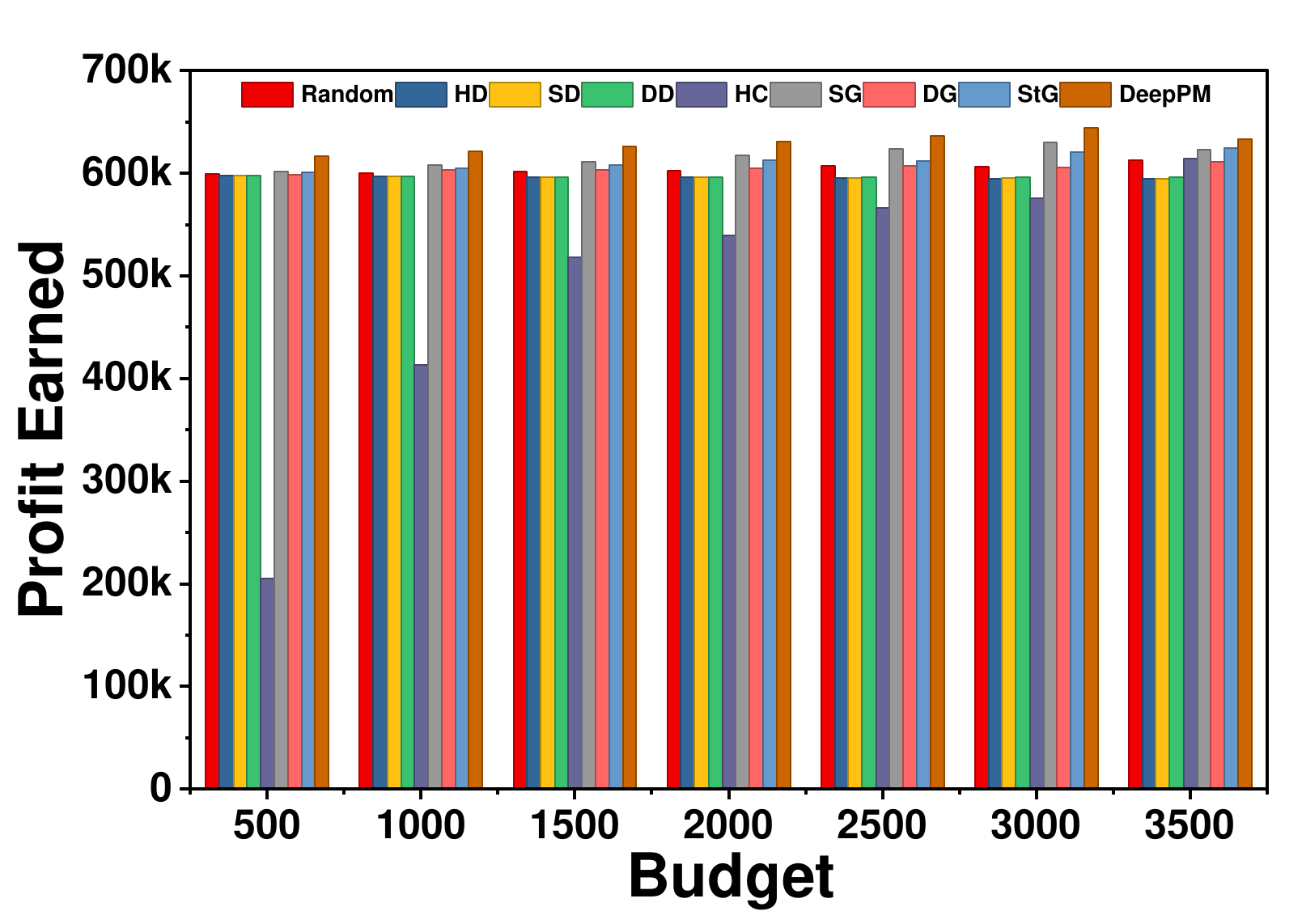}
    \caption{Euemail}
\end{subfigure}
\hfill
\begin{subfigure}{0.32\textwidth}
    \centering
    \includegraphics[width=\linewidth]{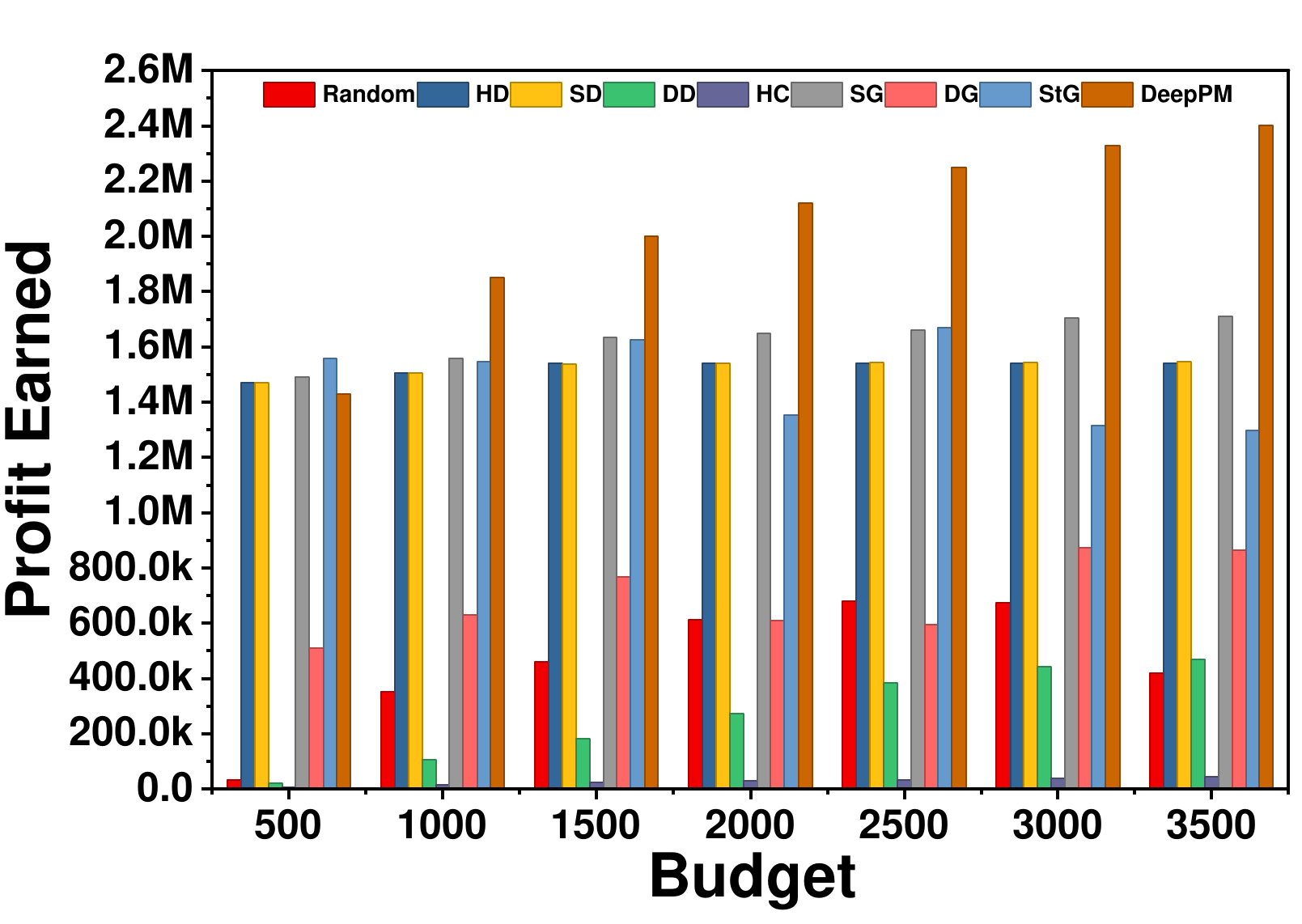}
    \caption{Facebook}
\end{subfigure}
\hfill
\begin{subfigure}{0.32\textwidth}
    \centering
    \includegraphics[width=\linewidth]{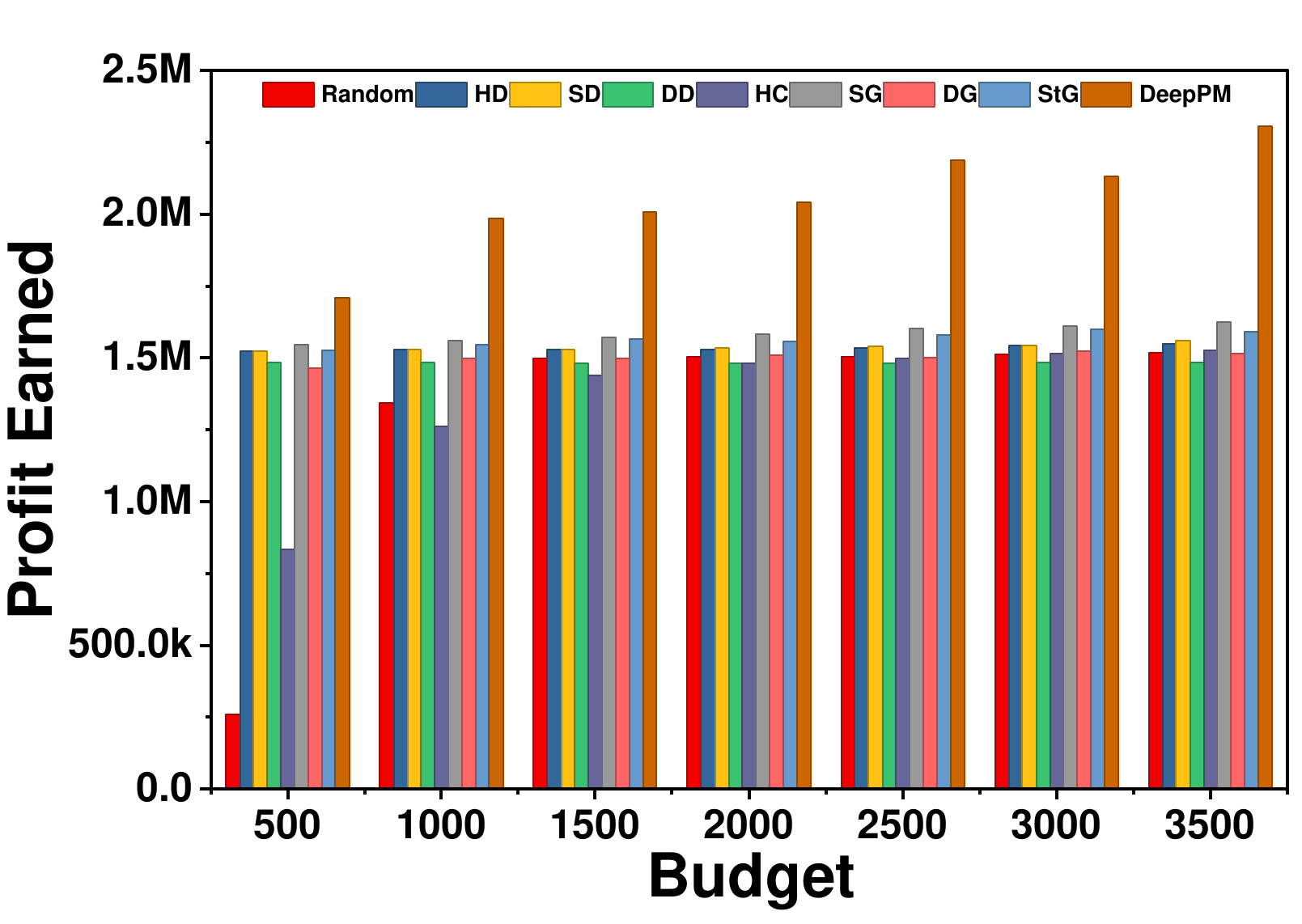}
    \caption{Wikivote}
\end{subfigure}

\caption{Plots showing Budget Vs. Profit Earned for Trivalency (a)-(c) and Uniform (d)-(f) Probability Setting}
\label{fig:profit}
\end{figure}

In this section, we examine how budget affects profit. Increasing the budget generally allows for the selection of more seeds, which in turn raises profit. In Fig. \ref{fig:profit}(a)-(c) and Fig. \ref{fig:profit}(d)-(f), profits rise with budget in both trivalency and uniform settings, respectively. Each algorithm spends its budget differently: \textbf{Random} selects nodes blindly; \textbf{HD} concentrates on high-degree nodes with overlapping influence, yielding only slight increases; \textbf{SD} avoids already-influenced neighbors but remains overly local, causing irregular profit gains; \textbf{DD} benefits steadily without oversaturating neighborhoods; \textbf{HC} suffers from heavy overlap, leading to declining profit as budget grows; \textbf{SG} and \textbf{StG} add seeds based on marginal gain (with \textbf{StG} adding randomization), producing strong improvements; and our \textbf{DeepPM}, which leverages the global structure learned by GCN, consistently increases profit as the budget expands. Firstly, we start with trivalency setting. For Euemail in Fig. \ref{fig:profit}(a), \textbf{DeepPM} earns 19\% more profit than \textbf{Random} and \textbf{DD}, 26\% more than \textbf{HD} and \textbf{SD}, and 13\% more than the strongest baseline, \textbf{SG}. For Facebook, in Fig. \ref{fig:profit}(b) \textbf{SG} and \textbf{StG} perform best among baselines, while \textbf{DeepPM} earns 21–58\% more than \textbf{SG} across budgets 500–3500.
For Wikivote (Fig. \ref{fig:profit}(c)), \textbf{SG} again leads the baselines, yet \textbf{DeepPM} achieves 60\% higher profit. Next is uniform setting. For Facebook (Fig. \ref{fig:profit}(e)), \textbf{SG} yields the maximum baseline profit, while \textbf{DeepPM} exceeds it by 18–40\%. For Wikivote (Fig. \ref{fig:profit}(c)), \textbf{SG} remains the best baseline, but \textbf{DeepPM} surpasses it by 10–42\%.

\subsubsection{RQ2. Impact of Budget on Seed Set Size}
\begin{figure}[h!]
\centering

\begin{subfigure}{0.32\textwidth}
    \centering
    \includegraphics[width=\linewidth]{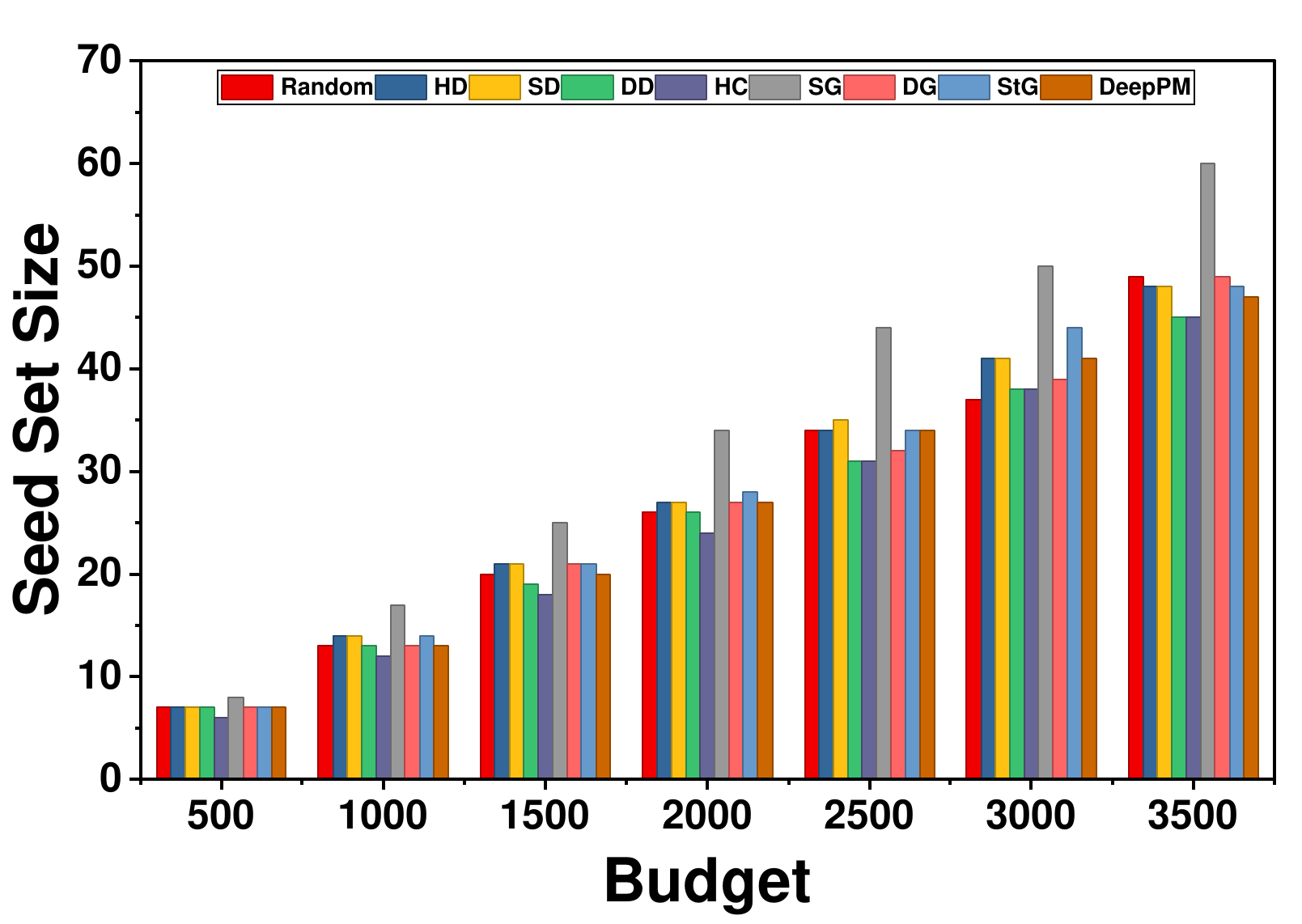}
    \caption{Euemail}
\end{subfigure}
\hfill
\begin{subfigure}{0.32\textwidth}
    \centering
    \includegraphics[width=\linewidth]{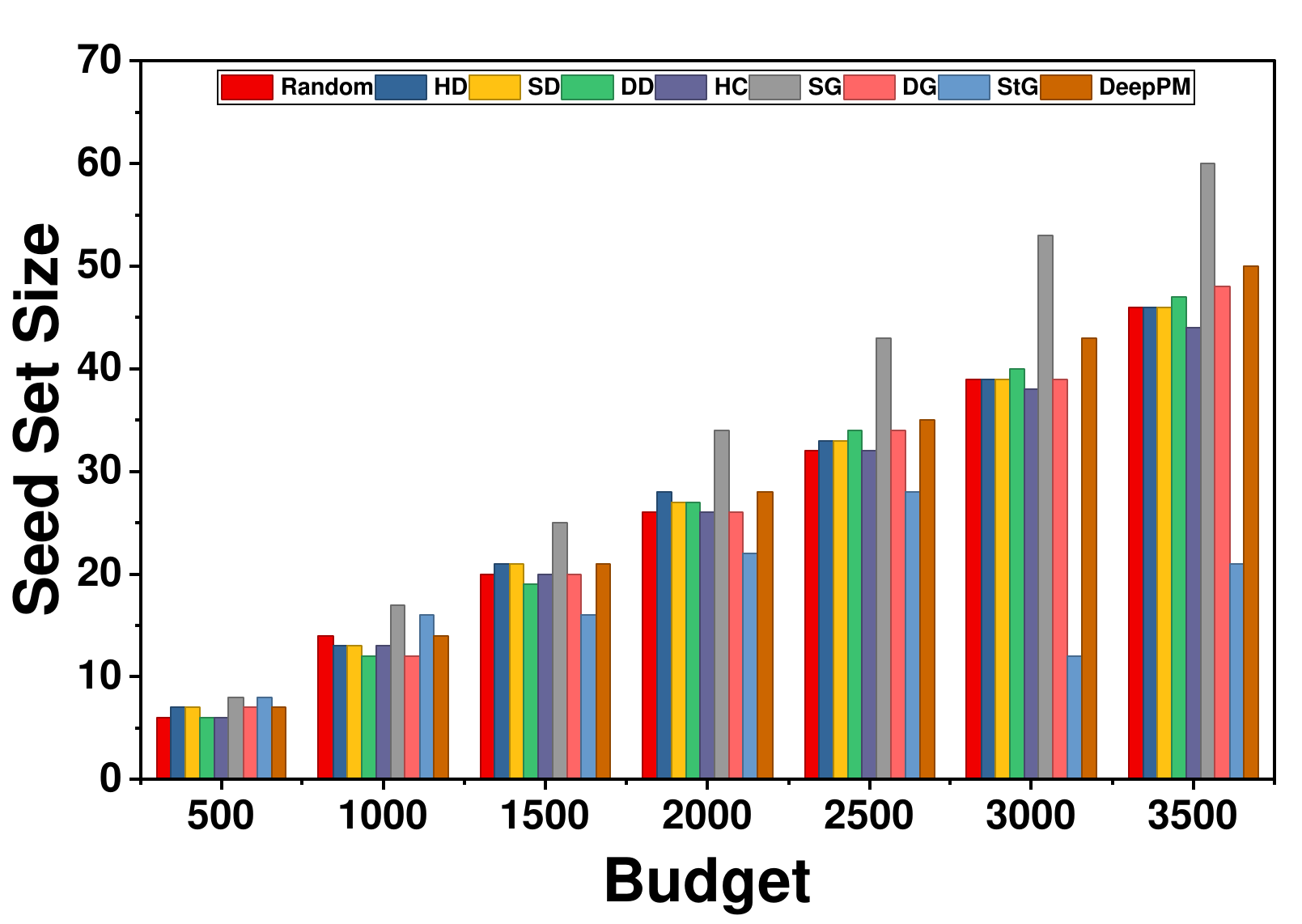}
    \caption{Facebook}
\end{subfigure}
\hfill
\begin{subfigure}{0.32\textwidth}
    \centering
    \includegraphics[width=\linewidth]{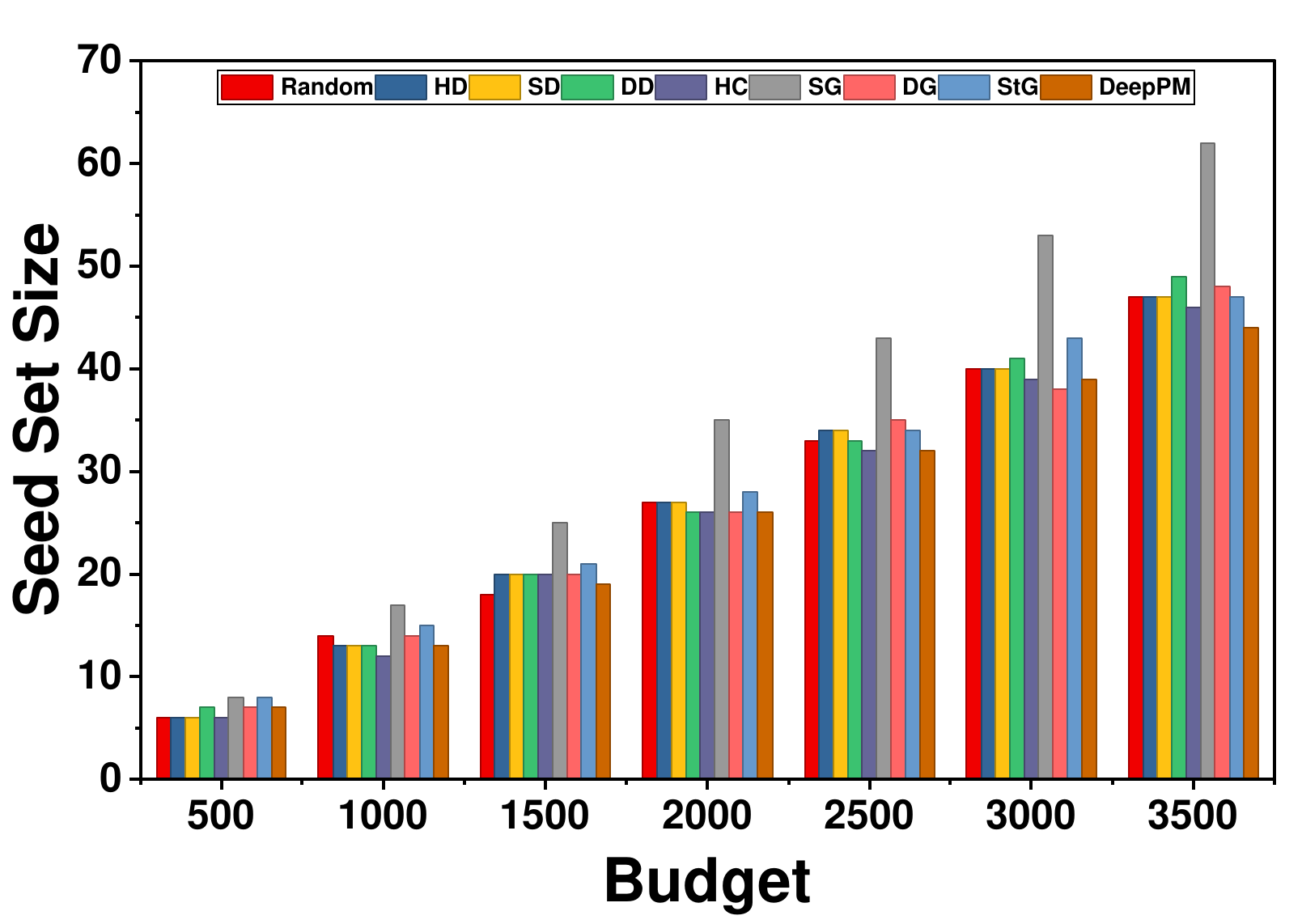}
    \caption{Wikivote}
\end{subfigure}

\vspace{8pt}

\begin{subfigure}{0.32\textwidth}
    \centering
    \includegraphics[width=\linewidth]{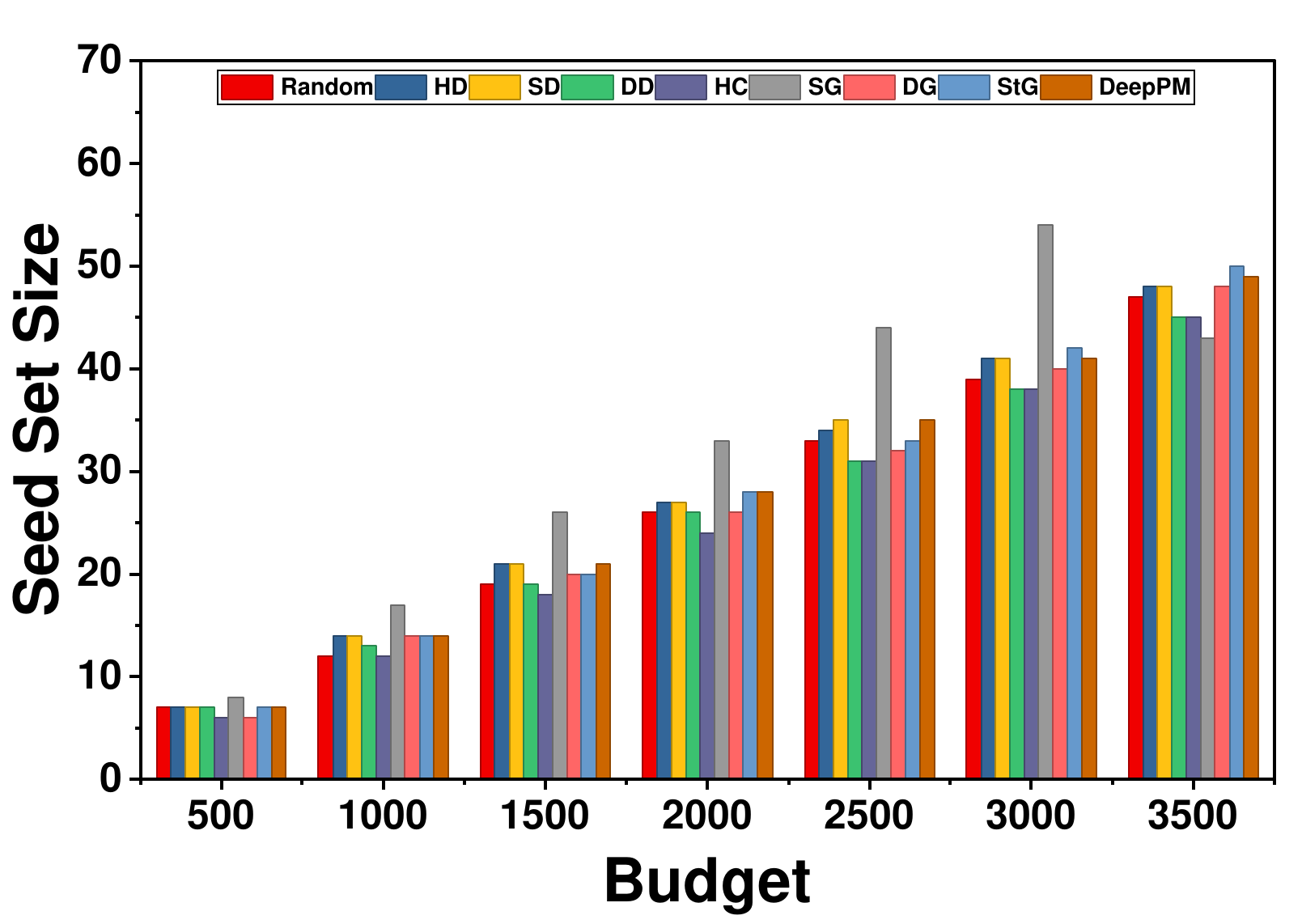}
    \caption{Euemail}
\end{subfigure}
\hfill
\begin{subfigure}{0.32\textwidth}
    \centering
    \includegraphics[width=\linewidth]{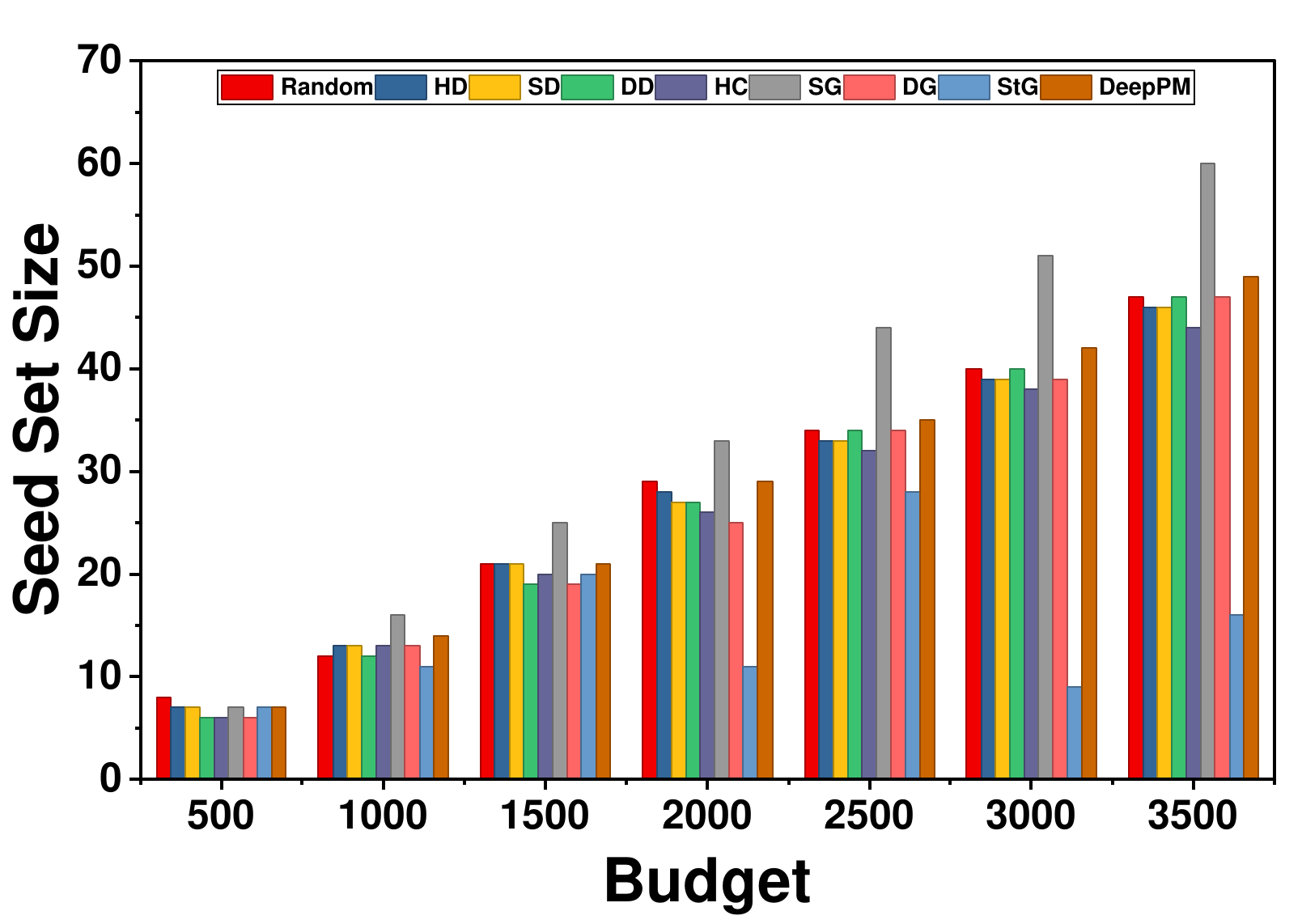}
    \caption{Facebook}
\end{subfigure}
\hfill
\begin{subfigure}{0.32\textwidth}
    \centering
    \includegraphics[width=\linewidth]{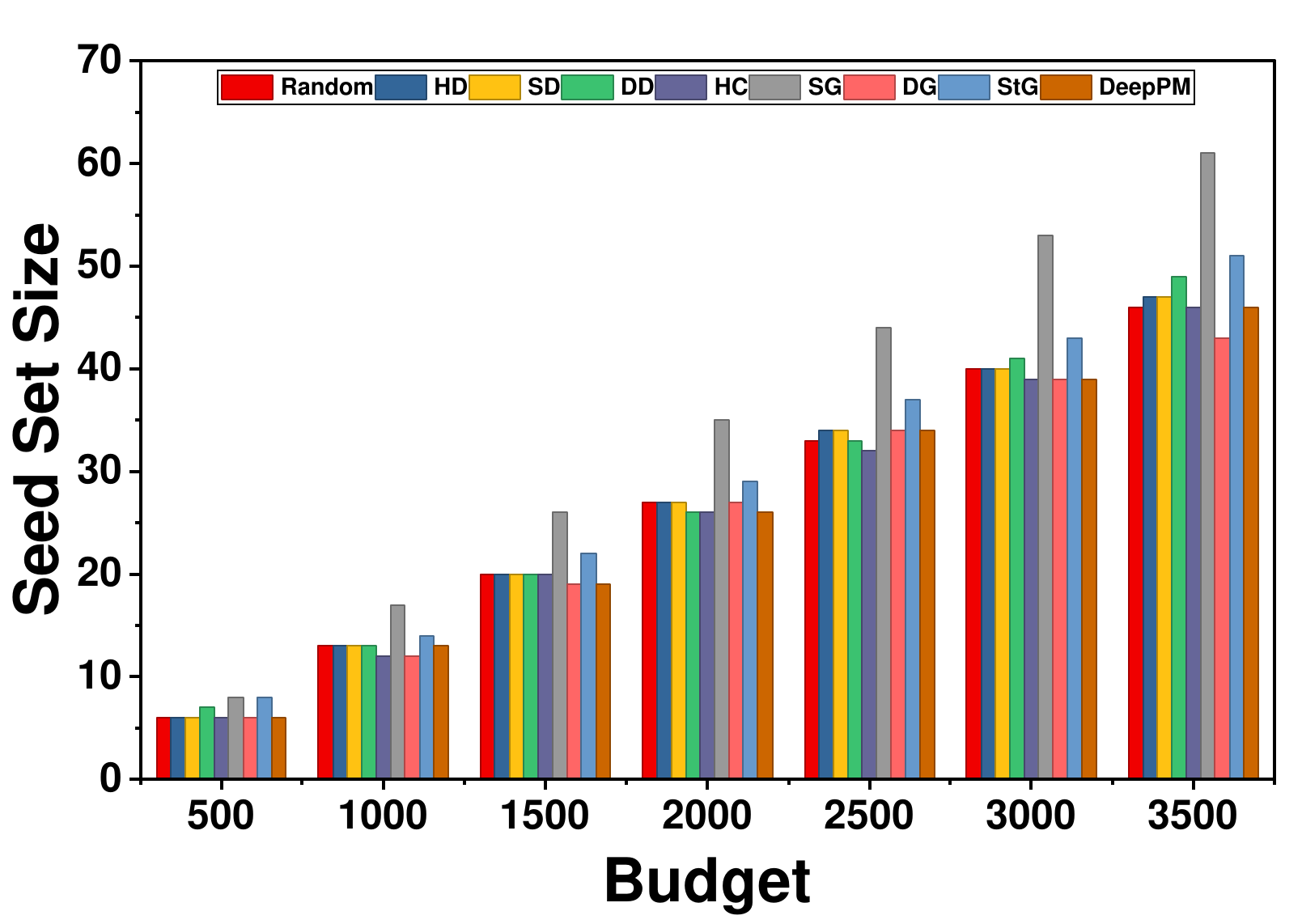}
    \caption{Wikivote}
\end{subfigure}

\caption{Plots showing Budget Vs. Seed Set Size for Trivalency (a)-(c) and Uniform (d)-(f) Probability Setting}
\label{fig:seedset}
\end{figure}

Here, we understand the impact of the budget on the seed set size. Given a budget X, we can have a seed set size ranging from the minimum possible size to the maximum size. The minimum size is the one that is calculated by dividing the maximum cost value node by the budget X, and the maximum size possible is calculated by dividing the minimum cost node by the budget X. We have seen that \textbf{SG} most often produces maximum profit among all baselines for datasets in both probability settings. In trivalency settings, we observe that the seed set size increases with the budget for all algorithms. For instance, in Fig. $\ref{fig:seedset}$(c), for \textbf{SG}, it has increased from $8$ (budget=500) to $62$ (budget=3500). Our \textbf{DeepPM} most often produces maximum profit compared to \textbf{SG}, but the size of the seed set remains moderate (less than that of \textbf{SG}), ranging from $7$ (budget=500) to $50$ (budget=3500). For the uniform setting shown in Fig. $\ref{fig:seedset}$(d)-(f), we have the same observations as in trivalency. The \textbf{SG} algorithm has a larger seed set size compared to the baselines, as well as our \textbf{DeepPM} approach. The maximum seed set size is 61 for \textbf{SG} with a budget of 3500, whereas for the same budget, \textbf{DeepPM} has a size of 46. Therefore, we observe that different algorithms exhibit distinct behavior under the same budget for selecting the seed set nodes. However, it is also true that the seed set size increases with an increasing budget.

\subsubsection{RQ3. Computational Time Requirement}
\begin{figure}[h!]
\centering

\begin{subfigure}{0.32\textwidth}
    \centering
    \includegraphics[width=\linewidth]{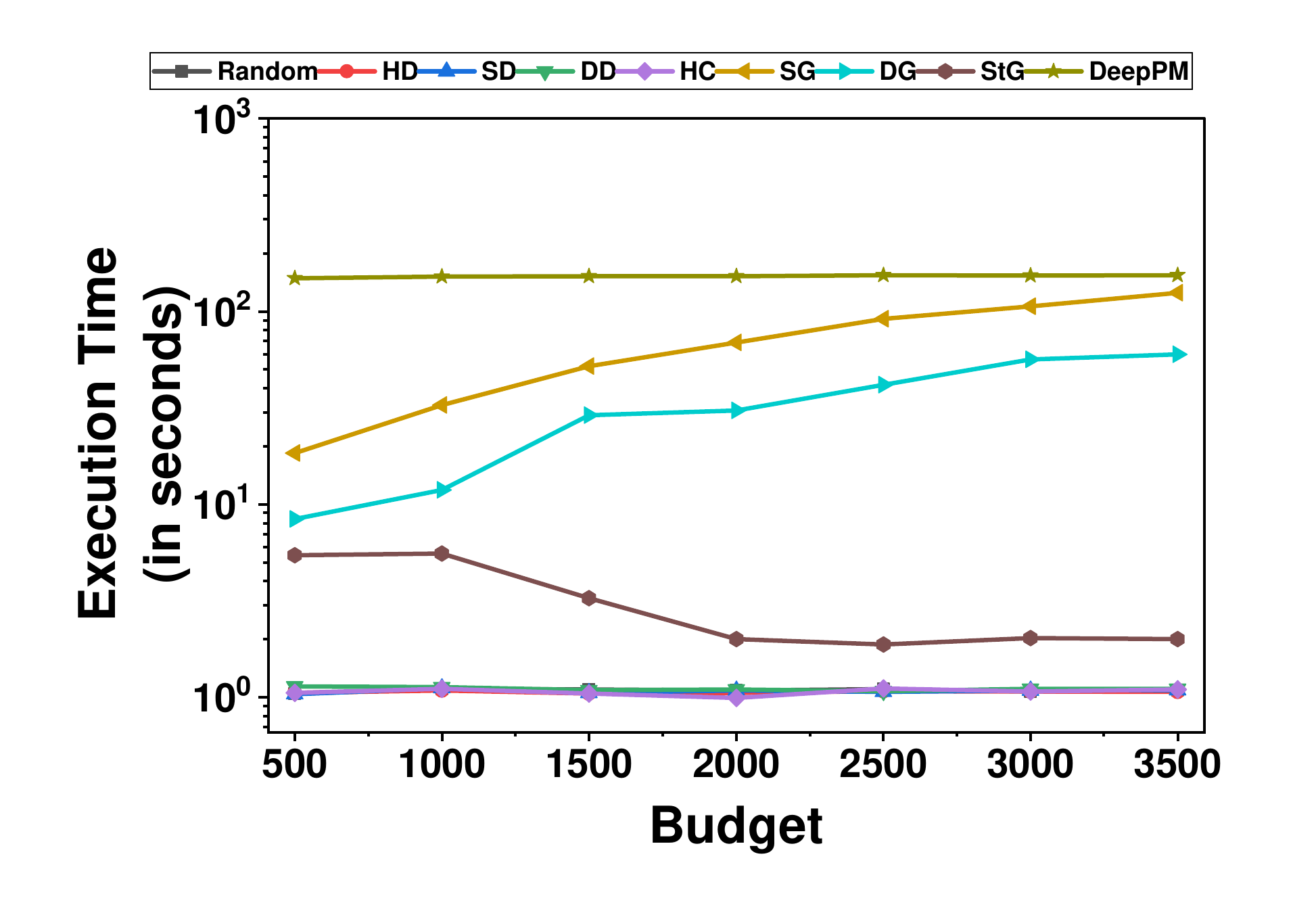}
    \caption{Euemail}
\end{subfigure}
\hfill
\begin{subfigure}{0.32\textwidth}
    \centering
    \includegraphics[width=\linewidth]{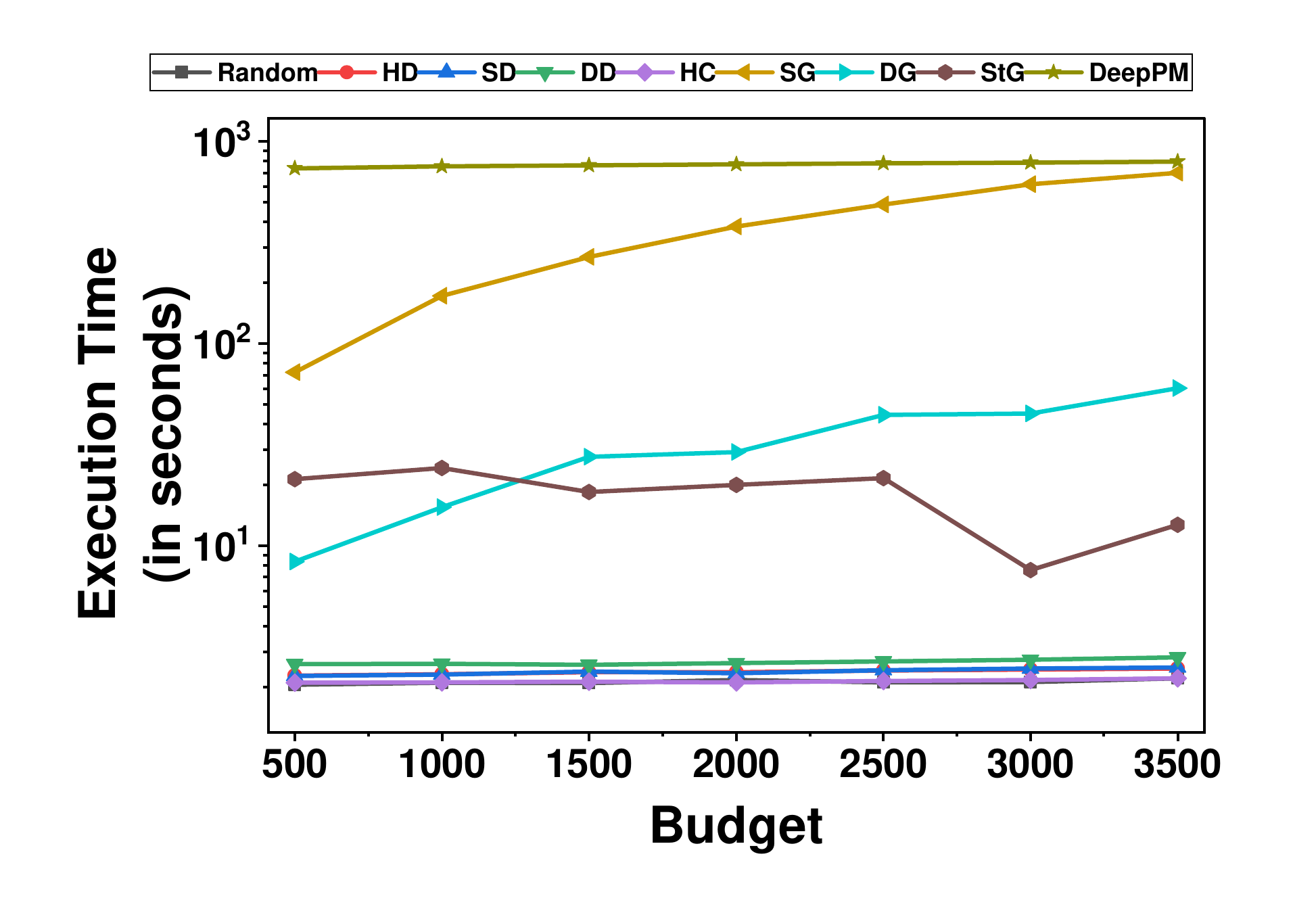}
    \caption{Facebook}
\end{subfigure}
\hfill
\begin{subfigure}{0.32\textwidth}
    \centering
    \includegraphics[width=\linewidth]{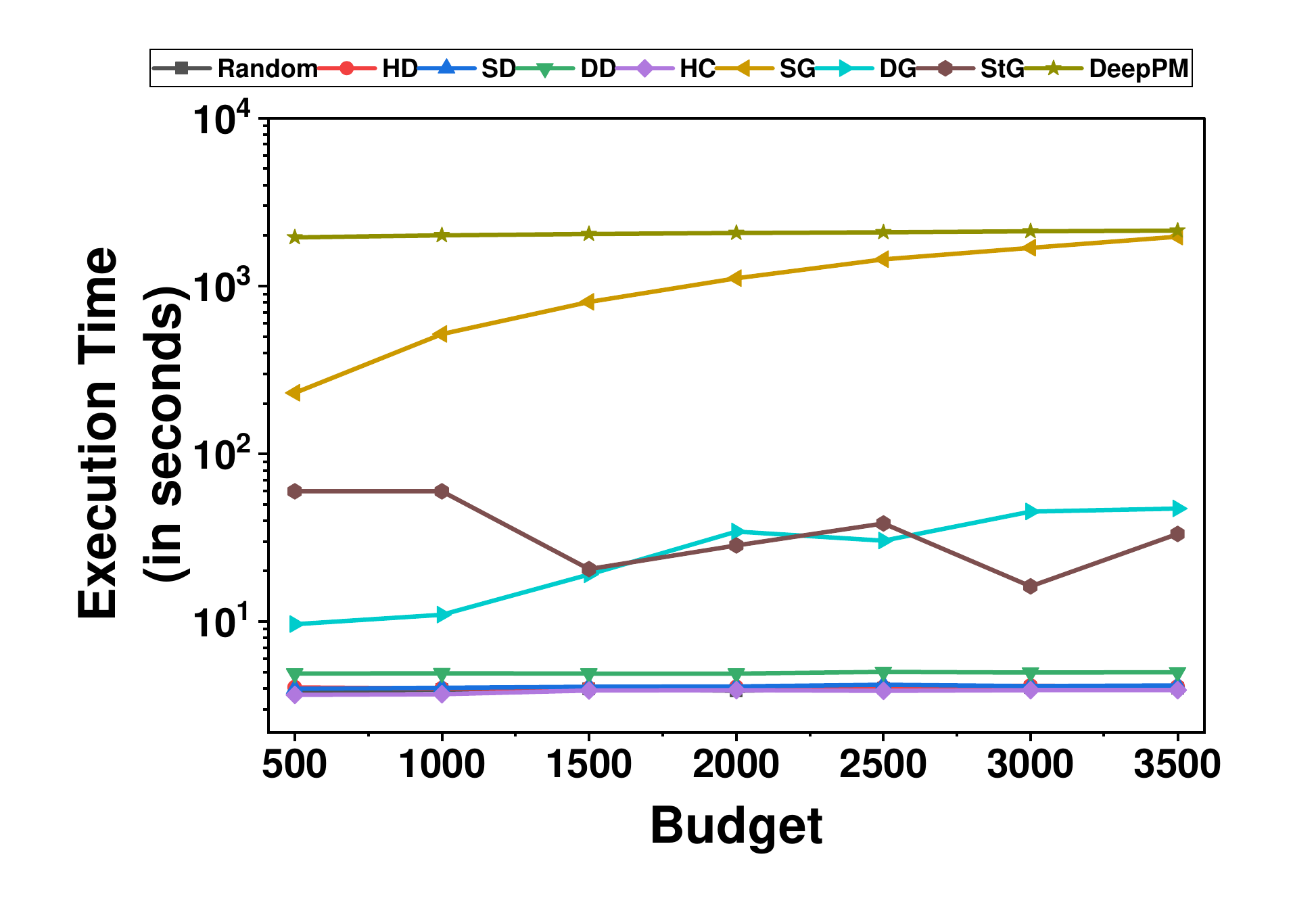}
    \caption{Wikivote}
\end{subfigure}

\vspace{8pt}

\begin{subfigure}{0.32\textwidth}
    \centering
    \includegraphics[width=\linewidth]{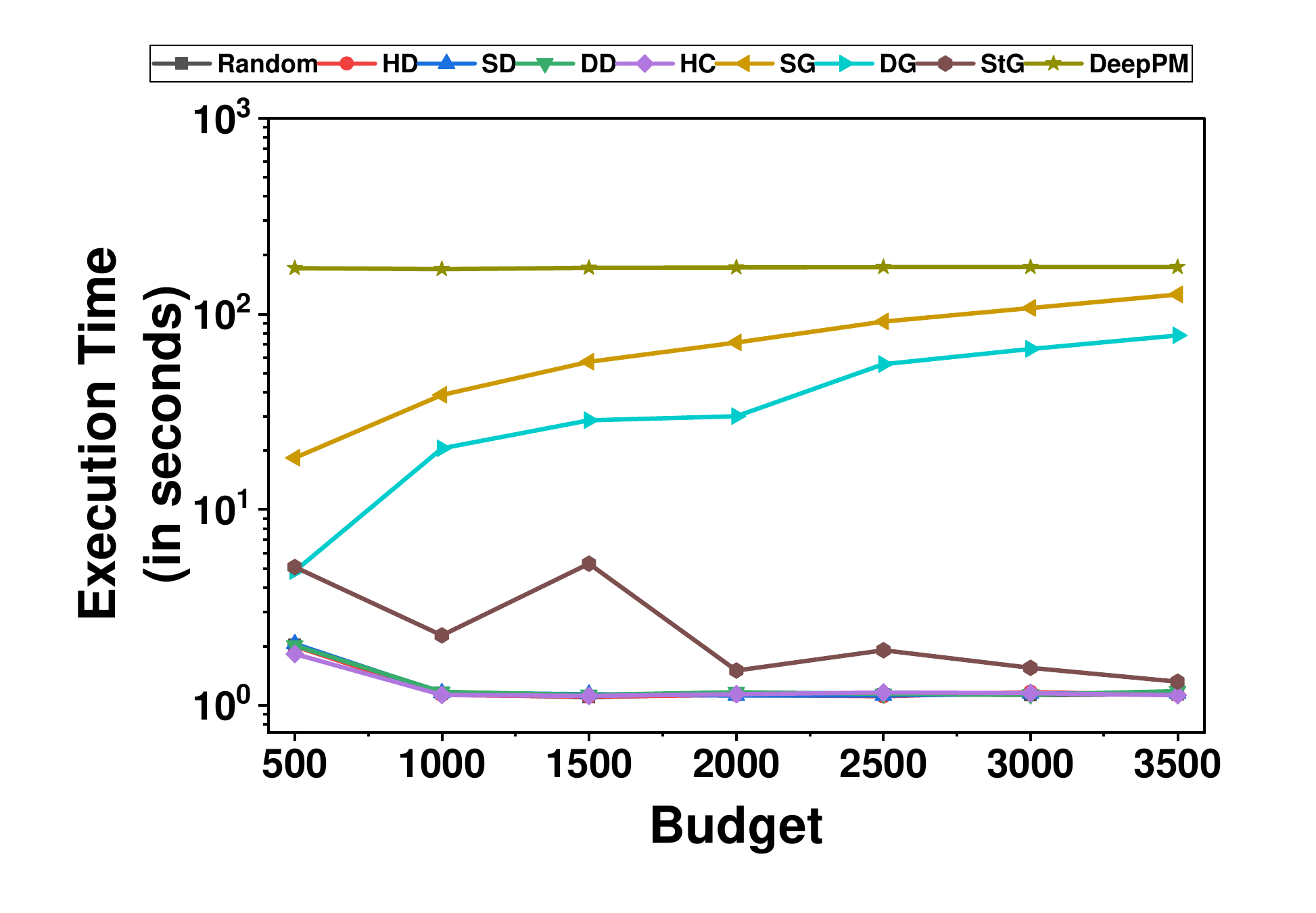}
    \caption{Euemail}
\end{subfigure}
\hfill
\begin{subfigure}{0.32\textwidth}
    \centering
    \includegraphics[width=\linewidth]{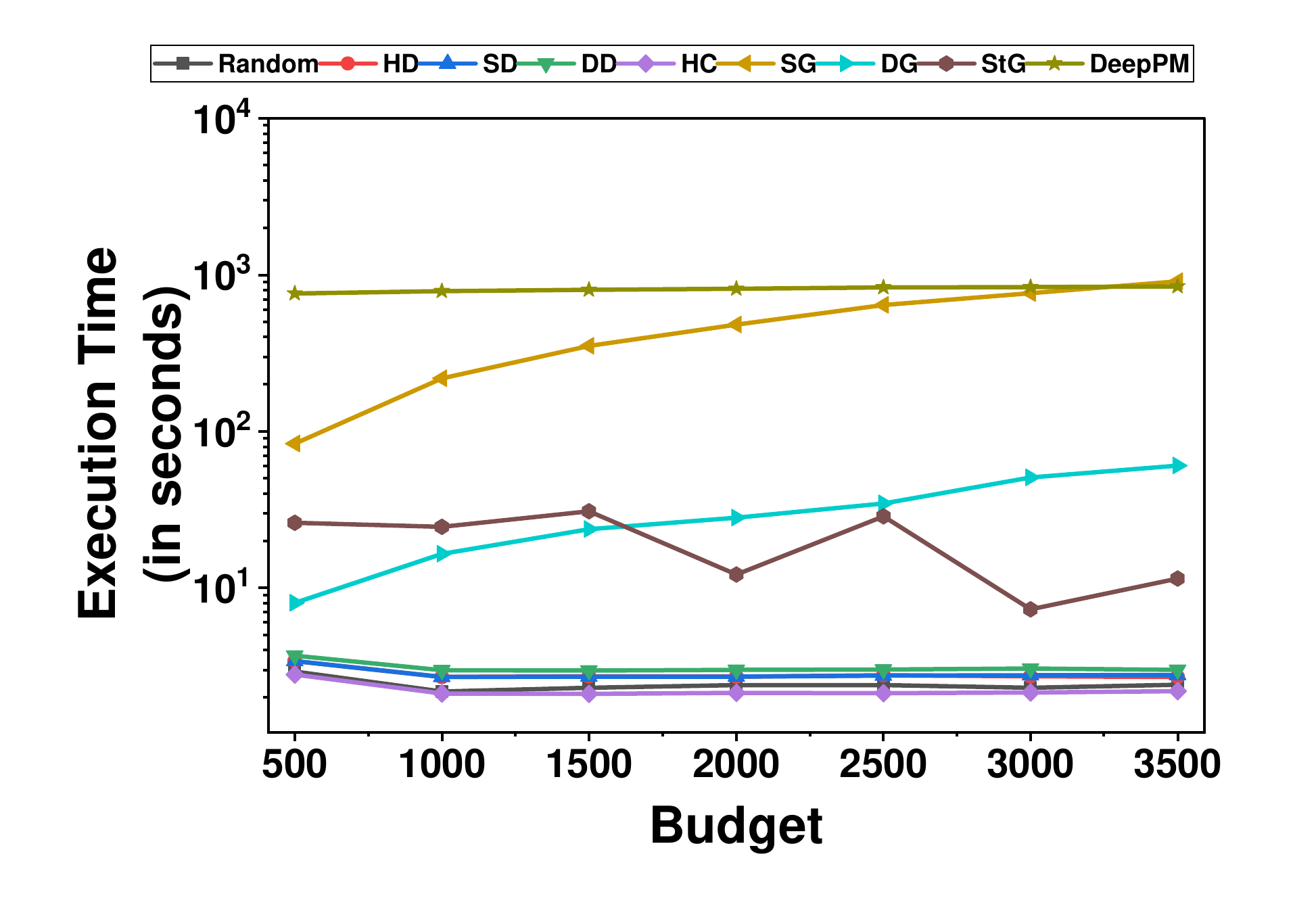}
    \caption{Facebook}
\end{subfigure}
\hfill
\begin{subfigure}{0.32\textwidth}
    \centering
    \includegraphics[width=\linewidth]{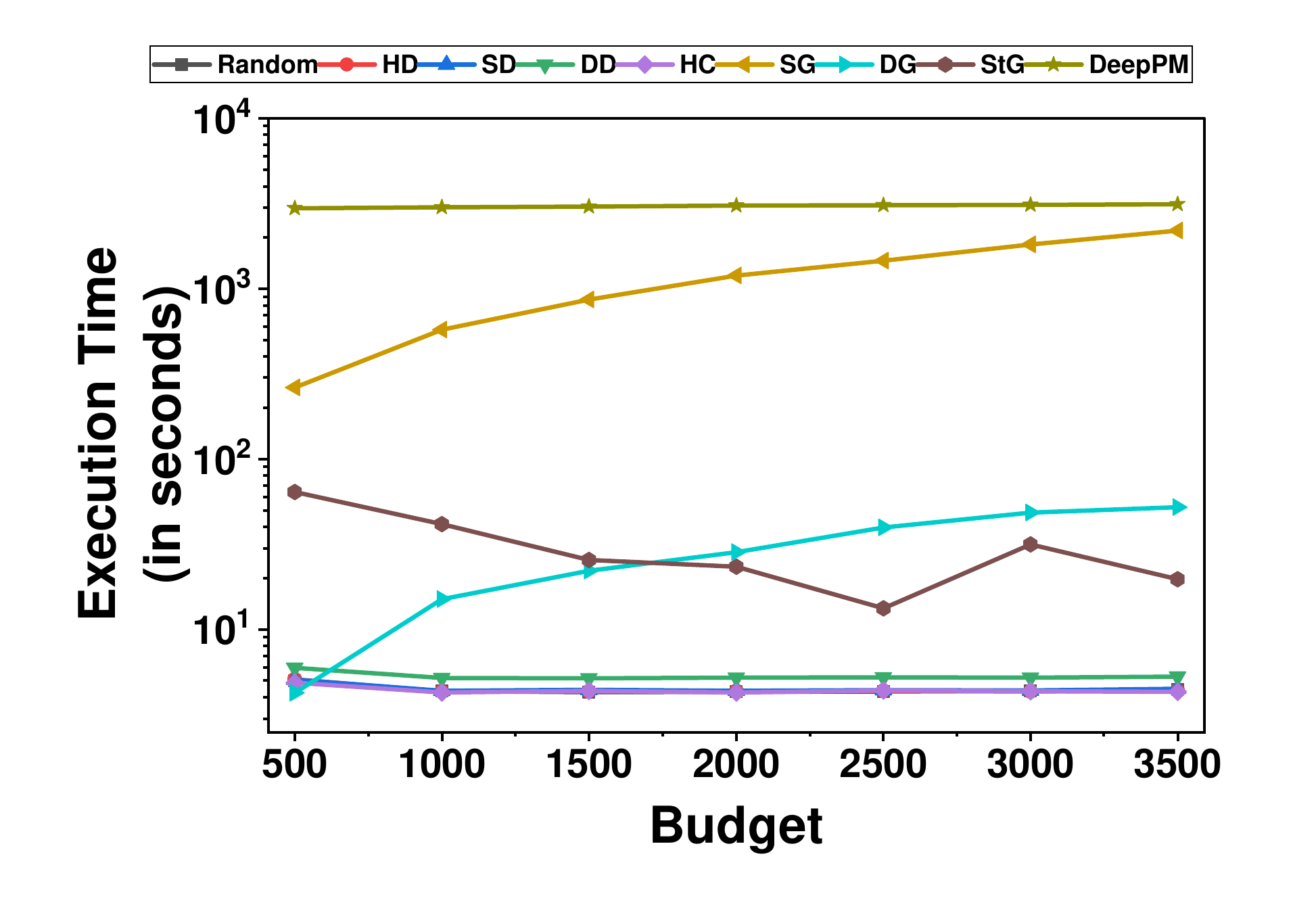}
    \caption{Wikivote}
\end{subfigure}

\caption{Plots showing Budget Vs. Execution Time (in seconds) for Trivalency (a)-(c) and Uniform (d)-(f) Probability Setting}
\label{fig:time}
\end{figure}

The computational time requirement of the baselines ranges from a few seconds to a few minutes for all datasets in both probability settings. In Fig. $\ref{fig:time}$(a)-(c), for trivalency probability, we see that the computational time of \textbf{SG} is the highest among all the baselines; however, \textbf{DeepPM} has even greater time requirements than \textbf{SG}. The \textbf{DeepPM} algorithm includes the time required for the computation of cascading, training, evaluation, and inferencing. We observe similar computational time trends for the uniform probability setting in Fig. $\ref{fig:time}$(d)-(f).

\section{Concluding Remarks} \label{Sec:Conclusion}
In this paper, we have studied the Profit Maximization Problem and proposed a deep learning-based solution approach. The proposed model has been elaborated in detail and experimentally tested using real-world social network datasets. We observe that the proposed solution approach yields a higher profit compared to many existing methods. Our future study on this problem will focus on developing learning-based solutions for large graph datasets.
%
%
\bibliographystyle{splncs04}
\bibliography{LaTeX_Template/sn-bibliography}

@inproceedings{yin2017local,
  title={Local higher-order graph clustering},
  author={Yin, Hao and Benson, Austin R and Leskovec, Jure and Gleich, David F},
  booktitle={Proceedings of the 23rd ACM SIGKDD international conference on knowledge discovery and data mining},
  pages={555--564},
  year={2017}
}

@inproceedings{leskovec2010signed,
  title={Signed networks in social media},
  author={Leskovec, Jure and Huttenlocher, Daniel and Kleinberg, Jon},
  booktitle={Proceedings of the SIGCHI conference on human factors in computing systems},
  pages={1361--1370},
  year={2010}
}

@inproceedings{leskovec2010predicting,
  title={Predicting positive and negative links in online social networks},
  author={Leskovec, Jure and Huttenlocher, Daniel and Kleinberg, Jon},
  booktitle={Proceedings of the 19th international conference on World wide web},
  pages={641--650},
  year={2010}
}

@inproceedings{chen2009efficient,
  title={Efficient influence maximization in social networks},
  author={Chen, Wei and Wang, Yajun and Yang, Siyu},
  booktitle={Proceedings of the 15th ACM SIGKDD international conference on Knowledge discovery and data mining},
  pages={199--208},
  year={2009}
}

@inproceedings{NIPS2012_7a614fd0,
 author = {Leskovec, Jure and Mcauley, Julian},
 booktitle = {Advances in Neural Information Processing Systems},
 editor = {F. Pereira and C.J. Burges and L. Bottou and K.Q. Weinberger},
 pages = {},
 publisher = {Curran Associates, Inc.},
 title = {Learning to Discover Social Circles in Ego Networks},
 volume = {25},
 year = {2012}
}

@ARTICLE{8241389,
  author={Tang, Jing and Tang, Xueyan and Yuan, Junsong},
  journal={IEEE Transactions on Knowledge and Data Engineering}, 
  title={Profit Maximization for Viral Marketing in Online Social Networks: Algorithms and Analysis}, 
  year={2018},
  volume={30},
  number={6},
  pages={1095-1108}
 }

@article{10.1007/s10878-021-00774-6,
author = {Du, Liman and Chen, Shengminjie and Gao, Suixiang and Yang, Wenguo},
title = {Nonsubmodular constrained profit maximization from increment perspective},
year = {2022},
issue_date = {Nov 2022},
publisher = {Springer-Verlag},
address = {Berlin, Heidelberg},
volume = {44},
number = {4},
issn = {1382-6905},
journal = {J. Comb. Optim.},
month = nov,
pages = {2598–2625},
numpages = {28}
}

@article{SHI202112,
title = {Profit maximization for competitive social advertising},
journal = {Theoretical Computer Science},
volume = {868},
pages = {12-29},
year = {2021},
issn = {0304-3975},
author = {Qihao Shi and Can Wang and Deshi Ye and Jiawei Chen and Sheng Zhou and Yan Feng and Chun Chen and Yanhao Huang}
}

@INPROCEEDINGS{7524470,
  author={Zhang, Huiyuan and Zhang, Huiling and Kuhnle, Alan and Thai, My T.},
  booktitle={IEEE INFOCOM 2016 - The 35th Annual IEEE International Conference on Computer Communications}, 
  title={Profit maximization for multiple products in online social networks}, 
  year={2016},
  volume={},
  number={},
  pages={1-9}}

@article{CHEN202036,
title = {A random algorithm for profit maximization in online social networks},
journal = {Theoretical Computer Science},
volume = {803},
pages = {36-47},
year = {2020},
issn = {0304-3975},
author = {Tiantian Chen and Bin Liu and Wenjing Liu and Qizhi Fang and Jing Yuan and Weili Wu}
}

@INPROCEEDINGS{8485904,
  author={Zhu, Yuqing and Li, Deying},
  booktitle={IEEE INFOCOM 2018 - IEEE Conference on Computer Communications}, 
  title={Host Profit Maximization for Competitive Viral Marketing in Billion-Scale Networks}, 
  year={2018},
  volume={},
  number={},
  pages={1160-1168}}

@article{XU201213009,
title = {Identifying valuable customers on social networking sites for profit maximization},
journal = {Expert Systems with Applications},
volume = {39},
number = {17},
pages = {13009-13018},
year = {2012},
issn = {0957-4174},
author = {Kaiquan Xu and Jiexun Li and Yuxia Song}
}

@ARTICLE{8630956,
  author={Yang, Wenguo and Yuan, Jing and Wu, Weili and Ma, Jianmin and Du, Ding-Zhu},
  journal={IEEE Transactions on Computational Social Systems}, 
  title={Maximizing Activity Profit in Social Networks}, 
  year={2019},
  volume={6},
  number={1},
  pages={117-126}}

@InProceedings{10.1007/978-3-030-34980-6_13,
author="Zhu, Jianming
and Ghosh, Smita
and Wu, Weili
and Gao, Chuangen",
editor="Tagarelli, Andrea
and Tong, Hanghang",
title="Profit Maximization Under Group Influence Model in Social Networks",
booktitle="Computational Data and Social Networks",
year="2019",
publisher="Springer International Publishing",
address="Cham",
pages="108--119",
isbn="978-3-030-34980-6"
}

@article{Zhu2013InfluenceAP,
  title={Influence and Profit: Two Sides of the Coin},
  author={Yuqing Zhu and Zaixin Lu and Yuanjun Bi and Weili Wu and Yiwei Jiang and Deying Li},
  journal={2013 IEEE 13th International Conference on Data Mining},
  year={2013},
  pages={1301-1306}}
\end{document}